\pdfoutput=1

\documentclass[11pt,a4paper]{article}
\usepackage{jheppub}
\usepackage{mciteplus}
\usepackage{hyperref}
\usepackage{url}
\usepackage{amssymb}
\usepackage{amsthm}
\usepackage{amstext}
\usepackage{amsmath}
\usepackage{amsfonts}
\usepackage[autostyle=true]{csquotes}

\usepackage{todonotes}

\usepackage{dsfont}
\usepackage{cancel}
\usepackage{wrapfig}
\usepackage{graphicx}
\usepackage{pstricks}
\usepackage{color}

\usepackage{enumitem}

\usepackage{pgf,tikz}
\usetikzlibrary{calc,shapes,automata,petri}
\tikzset{%
  transition/.style={rectangle,minimum size=6mm,draw},
  place/.style={circle,minimum size=6mm,draw},
  database/.style={
    minimum width=2cm,minimum height=1cm,cylinder,
    shape border rotate=90,aspect=0.2,draw
  }
}

\newlength{\lw}
\newlength{\st}
\newlength{\qs}
\newlength{\nd}
\tikzstyle{every picture}+=[auto]
\tikzstyle{every picture}+=[bend angle=10]
\tikzstyle{every picture}+=[join=round]
\tikzstyle{every picture}+=[cap=butt]
\tikzstyle{every picture}+=[line width=\lw]
\tikzstyle{every picture}+=[double distance=2\lw]
\tikzstyle{every picture}+=[shorten >=\st]
\tikzstyle{every picture}+=[node distance=\nd]
\tikzstyle{every loop}=[->,shorten >=\st]
\tikzstyle{place}=[circle,draw,minimum size=\qs]
\tikzstyle{transition}=[rectangle,draw,minimum size=\qs]
\tikzstyle{invisible}=[draw=none,inner sep=0pt,minimum height=0pt]
\newenvironment{petrinet}[1][]{\begin{center}\begin{tikzpicture}}{\end{tikzpicture}\end{center}}

\usepackage{xparse}

\def\d{{\rm d}}

\NewDocumentCommand\Tensor{mmggg}{
#1 \otimes #2 \IfNoValueTF{#3}{}{\otimes #3} \IfNoValueTF{#4}{}{\otimes #4} \IfNoValueTF{#5}{}{\otimes #5}
}

\theoremstyle{plain}

\numberwithin{equation}{section}

\title{Integration-by-parts reductions of Feynman integrals using
  Singular and GPI-Space}
\author[a, b]{Dominik~Bendle}
\author[a]{Janko~B{\"o}hm}
\author[a]{Wolfram~Decker}
\author[c]{Alessandro~Georgoudis}
\author[b]{Franz-Josef~Pfreundt}
\author[b]{Mirko~Rahn}
\author[d]{Pascal~Wasser}
\author[e,f]{Yang~Zhang}
\affiliation[a]{Department of Mathematics, Technische Universit\"at
  Kaiserslautern, 67663 Kaiserslautern, Germany}
\affiliation[b]{Fraunhofer Institute for Industrial Mathematics ITWM,
  Fraunhofer-Platz 1, 67663 Kaiserslautern, Germany}
\affiliation[c]{Department of Physics and Astronomy, Uppsala
  University, SE-75108 Uppsala, Sweden}
\affiliation[d]{PRISMA+ Cluster of Excellence, Johannes Gutenberg University, D-55128 Mainz, Germany}
\affiliation[e]{Interdisciplinary Center for Theoretical Study,
  University of Science and Technology of China, Hefei, Anhui 230026,
  China}
\affiliation[f]{Max-Planck-Institut f\"ur Physik, Werner-Heisenberg-Institut, D-80805 M\"unchen, Germany}

\emailAdd{bendle@rhrk.uni-kl.de}
\emailAdd{boehm@mathematik.uni-kl.de}
\emailAdd{decker@mathematik.uni-kl.de}
\emailAdd{Alessandro.Georgoudis@physics.uu.se}
\emailAdd{franz-josef.pfreundt@itwm.fhg.de}
\emailAdd{mirko.rahn@itwm.fhg.de}
\emailAdd{wasserp@uni-mainz.de}
\emailAdd{yzhang@mpp.mpg.de}

\abstract{We introduce an algebro-geometrically motived integration-by-parts (IBP)
  reduction method for multi-loop and multi-scale Feynman
  integrals, using a framework for massively parallel computations in computer algebra. This framework combines the computer algebra system \textsc{Singular} with the workflow management system
  \textsc{GPI-Space}, which are being developed at the TU Kaiserslautern and the Fraunhofer Institute for Industrial Mathematics (ITWM), respectively. In our approach, the IBP relations are first trimmed by modern tools from
  computational algebraic geometry  and then solved by sparse linear algebra
  and our new interpolation method. Modelled in terms of Petri nets, these steps are efficiently automatized and automatically parallelized by \textsc{GPI-Space}. We demonstrate
  the potential of our method at the nontrivial example of reducing two-loop five-point
 nonplanar double-pentagon integrals. We also use \textsc{GPI-Space} to
  convert the basis of IBP reductions, and discuss the possible simplification
  of IBP coefficients in a uniformly transcendental basis.}


\begin{document}

\begin{flushright}\begin{tabular}{r}
MITP/19-055 \\
MPP-2019-164\\
UUITP-30/19\\
USTC-ICTS-19-20
\end{tabular}\end{flushright}
\vspace{-14.1mm}
\maketitle

\section{Introduction}
With the success of Large Hadron Collider (LHC) Run II and the upcoming
LHC run III, high precision background computation, especially
next-to-next-to-leading-order (NNLO) scattering computation, is crucial
for the interpretation of experimental results. In recent years, great progress has been made in multi-loop scattering amplitude
calculations, for instance, in the case of $2 \rightarrow 3$ processes
\cite{Badger:2013gxa,Gehrmann:2015bfy,Badger:2017jhb,Abreu:2017hqn,Abreu:2018aqd,Abreu:2018jgq,Boels:2018nrr,Gehrmann:2018yef,
Badger:2018enw,Abreu:2018zmy,Chicherin:2018yne,Chicherin:2019xeg,Abreu:2019rpt,Abreu:2019odu,
Hartanto:2019uvl}. 
The progress is due to modern developments of scattering amplitudes,
like the integrand construction method
\cite{Zhang:2012ce,Mastrolia:2012an}, canonical integrals
\cite{Henn:2013pwa,Henn:2014qga}, numeric unitarity
\cite{Ita:2015tya,Abreu:2017xsl},  bootstrap methods
\cite{Dixon:2011pw,Dixon:2013eka,Dixon:2014iba,Caron-Huot:2016owq,Dixon:2015iva,Dixon:2016nkn,Chicherin:2017dob,Caron-Huot:2019vjl},
reconstruction using finite fields
\cite{vonManteuffel:2014ixa,Peraro:2016wsq,Klappert:2019emp, Peraro:2019svx} and new
ideas in the integration-by-parts  (IBP)  reduction. The latter is the main topic to be discussed in
this paper. 

Frequently, when computing scattering amplitudes, IBP reduction is
a crucial and bottleneck step. It is a fundamental tool for both the reduction of integrals to master integrals (MIs),  and for computing the master integrals themselves using the differential equation method.
IBP relations (IBPs) are derived from integrating a total derivative \cite{CHETYRKIN1981159},
\begin{equation}
0=\int \frac{\d^D \ell_1}{\mathrm{i} \pi^{D/2}} \ldots \frac{\d^D \ell_L}{\mathrm{i} \pi^{D/2}}
\sum_{j=1}^L \frac{\partial}{\partial \ell_j^\mu}
\frac{v_j^\mu \hspace{0.5mm} }{D_1^{\nu_1} \cdots D_m^{\nu_m}} \,, \label{eq:IBP_schematic}
\end{equation}
where the $v_i^\mu$ are polynomials in the loop momenta $\ell_i$, the $D_i$
are the inverse propagators, and $D$ is the spacetime dimension. 

The standard approach to obtain IBP reductions, by which we are able
to express an integral as a linear combination of a finite number of
MIs, is to generate sufficiently many IBP relations, and then use the
Laporta algorithm \cite{Laporta:2001dd} to solve the associated linear
system. The algorithm works by imposing an ordering on the different
integral families and solving recursively. There exist multiple
public and private implementations of this approach
\cite{Smirnov:2008iw,Smirnov:2014hma, Smirnov:2019qkx,
  Maierhoefer:2017hyi, Maierhofer:2018gpa, vonManteuffel:2012np,Klappert:2019emp},
which usually generates a large linear system to be solved.

In the case of a system of IBPs which does not have double propagators \cite{Gluza:2010ws,Schabinger:2011dz,Larsen:2015ped}, however,
we obtain a much smaller linear system. The IBPs without double
propagators are physically related to dual conformal symmetries \cite{Bern:2017gdk}. A significant simplification can be made by using unitarity methods, where
by considering a spanning set of generating cuts it is possible to
reduce the size of the IBP system. This requires prior knowledge
of a basis of MIs. Such a basis can be obtained by running the Laporta algorithm
with constant kinematics, or by using specialized
programs such as {\sc Mint} \cite{Lee:2013hzt} or {\sc Azurite}
\cite{Georgoudis:2016wff}. (Note that the dimension of a basis of integrals 
can also be obtained by studying the parametric annihilators
\cite{Bitoun:2017nre}.) There is also the important technique \cite{Chawdhry:2018awn} of simultaneously
nullifying all master integrals except one, which often makes large-scale linear reductions feasable.

Besides the advances in purely analytical methods in recent years,
there has been a lot of work towards numerical implementations of the
generation of IBPs. The idea is to utilize either integer values or
finite-field values for the kinematical invariants \cite{vonManteuffel:2014ixa,Peraro:2016wsq,Smirnov:2019qkx}, depending on the difficulty of the problem, and then to
run the same reduction several times for reconstruction. This method has been very
successful in tackling difficult problems.  Furthermore, it is
possible to numerically generate and reduce the IBP relations, and, while skipping the IBP coefficient
reconstruction, directly carry out an amplitude reconstruction. (For examples,
see \cite{Badger:2018enw, Abreu:2018zmy, Abreu:2019rpt, Badger:2019djh}). 
In this paper, we in particular present
our own implementation of a semi-numeric rational
interpolation method, see Appendix \ref{sec:rational_func_interpolation} for more details.

Furthermore, new approaches were developed recently to obtain the reduction directly, without generating IBP relations from total derivatives. In \cite{Kosower:2018obg}, the direct
solution method was presented to derive recurrence relations for
Feynman integrals and solve them analytically with arbitrary numerator
degree. One very promising progress is based on the
intersection theory of differential forms in the Baikov representation
\cite{Mastrolia:2018uzb,Frellesvig:2019kgj,Frellesvig:2019uqt}. This
approach calculates the master integral coefficients from 
intersection numbers. There is also a very intuitive approach to
reduce Feynman integrals by considering the $\eta$ expansion of the
Feynman prescription \cite{Liu:2018dmc, Liu:2017jxz, Wang:2019mnn}. 
Using this approach,  the scaling of the reduction computation depends only linearly on the number of master
integrals. Furthermore, it is possible to determine two-loop planar
diagram IBP coefficients directly from the Baikov representation \cite{Kardos:2018uzy}.

In this paper, we present our new powerful IBP reduction method based
on:
\begin{enumerate}
\item Computational algebraic geometry. We apply the module
  intersection method from \cite{Boehm:2018fpv,Boehm:2017wjc}, modified by using
  a suitably chosen degree bound for the Gr\"obner basis computation, to efficiently
  generate a small IBP system, without double propagators (or IBPs with a given bound on the
  propagator exponents).
\item A modern framework of massively parallel computations in
  computer algebra which combines the computer algebra system
  \textsc{Singular} \cite{singular} with the workflow management
  system \textsc{GPI-Space} \cite{GPI}. We have completely automatized
  our approach and make our algorithms run automatically in parallel
  on high performance computing clusters. In this way, IBP results can
  be obtained in an efficient, reliable and scalable way.
  Our implementation can automatically determine the minimal number of
  points needed for interpolating the IBP coefficients, it can identify possible 
``bad'' points, add more points, if necessary, and interpolate the final result. 
\end{enumerate}

We demonstrate the power of our method by reducing the two-loop
five-point nonplanar double pentagon diagram {\it analytically}, up to
numerator degree $4$. This is a nontrivial test since the diagram
has a complicated topology and there are
five symbolic Mandelstam variables as well as the spacetime variable $D$. 

Furthermore, we start to look at the possible simplification of IBP
coefficients by converting the master integral basis. In this paper,
we test the conversion to a ``dlog'' basis \cite{WasserMSc}, a special case of the
canonical basis \cite{Henn:2014qga}. We find that for the double pentagon diagram above,
the size of the IBP coefficients reduces significantly  from the byte size $\sim 2.0$G in the Laporta basis to
$\sim 0.48$G in the dlog basis on disk, that is, by $76\%$.  The master integral basis conversion
computation is also automated by the \textsc{Singular}-\textsc{GPI-Space} framework.

Our paper is structured as follows. In Section \ref{sec:Mod_int}, we
present the general background on how to generate simple and trimmed
IBP systems using computational algebraic geometry and finite-field
methods, as well as the improvement on the algorithm in
\cite{Boehm:2018fpv}. In Section \ref{sec:gpising}, we give a short overview on how we use \textsc{Singular} in conjunction with \textsc{GPI-Space}. 
In Section \ref{sec IBP Petri}, we describe how to model our algorithm
in the \textsc{Singular}-\textsc{GPI-Space} framework, and discuss timings and scaling of the algorithm, focusing on the double pentagon diagram. This, in
particular, demonstrates the potential of the
\textsc{Singular}-\textsc{GPI-Space} framework for applications in
high-energy physics. In Section \ref{sec:IBP_conv}, we review the algorithmic computation
of a dlog basis which has uniform transcendental weight, and we comment on how to convert coefficients
from the Laporta basis to the dlog basis. In Section
\ref{sec:doublepentagon}, we study the working example of our
implementation, the double pentagon graph, in detail. We discuss the analytic IBP
reduction and the conversion of IBP coefficients to the dlog basis. Finally
we present a summary and conclusion of this paper.

The result of our IBP reductions can be downloaded from the following
links: Whereas

\begin{small}
\noindent \url{https://www.dropbox.com/s/1ubdhcyhe8e4pwy/IBPmatrix_Laporta_basis.tar.gz}\\
\end{small}
provides the IBP coefficients in the Laporta basis with the scale
$s_{12}=1$,

\begin{small}
\noindent \url{https://www.dropbox.com/s/e6t4evftkfo95pr/IBPmatrix_dlog_basis.tar.gz}\\
\end{small}
contains the IBP coefficients in the dlog basis with the scale
$s_{12}=1$. 

For the convenience of the reader, we also present the IBP
coefficients in the dlog basis with the full scale dependence:

\begin{small}
\noindent \url{https://www.dropbox.com/s/dnkr6h5t3vik2r0/IBPmatrix_dlog_basis_scaled.tar.gz}
\end{small}

We encourage researchers in the high energy community to send us IBP
reduction problems  (mailto: \href{mailto:alessandro.georgoudis@physics.uu.se}{alessandro.georgoudis@physics.uu.se}) for cutting-edge precision calculations and
the further sharpening of our new reduction method.

\section{The module intersection method reloaded}
\label{sec:Mod_int}
In this section, we present a refined version of the approach of using module
intersections to trim IBP systems. For the detailed account of the module
intersection IBP reduction method, we refer to \cite{Boehm:2018fpv}.

\subsection{Module intersection}
The Feynman integrals under consideration are labeled as
\begin{equation}
  I[n_1,\ldots, n_m] = \int \frac{\d^D l_1}{i \pi^{D/2}} \ldots
  \frac{\d^D l_L}{i \pi^{D/2}}  \frac{1}{D_1^{n_1} \cdots
    D_m^{n_m}}\,,
\label{Feynman_integral}
\end{equation}
where $L$ is the loop order and the $l_i$'s are the loop momenta. We have $E$ independent external vectors that we label as $p_1, ..., p_E$. We assume that the
Feynman integrals have been reduced on the integrand level, and set
$m=LE+L(L+1)/2$ which equals the number of scalar products in the
configuration.

For us it is convenient to use the Baikov representation
\cite{Baikov:1996rk,Lee:2013hzt} for IBP
reductions,
\begin{align}
I[n_1,\ldots, n_m]  \hspace{0.5mm}&=\hspace{0.5mm}  C_E^L \hspace{0.7mm} U^\frac{E-D+1}{2} \hspace{-1mm}
\int \d z_1 \cdots \d z_m P^\frac{D-L-E-1}{2} \frac{1}{z_1^{n_1} \cdots z_m^{n_m}}  \,.
\label{eq:Baikov_representation}
\end{align}
Here, $P$ is the Baikov polynomial, which can be written as a Gram determinant,
\begin{equation}
  P=\det G\left(
    \begin{array}{cccccc}
      l_1,& \ldots &l_L, &p_1, & \ldots & p_E \\
l_1,& \ldots &l_L, &p_1, & \ldots & p_E
    \end{array}
\right) \,.
\end{equation}
Moreover, $U$ and $C_E^L$ are the Gram determinant respectively constant factor below:
\begin{equation}
  U=\det G\left(
    \begin{array}{ccc}
     p_1, & \ldots & p_E \\
p_1, & \ldots & p_E
    \end{array}
\right) , \quad
C_E^L =J\frac{\pi^\frac{L-m}{2}}{\Gamma(\frac{D-E-L+1}{2})\ldots \Gamma(\frac{D-E}{2})}\,,
\end{equation}
where $J$ is a constant Jacobian. The factors $U$ and $C_E^L$ are
irrelevant for the IBP relations.

As in \cite{Lee:2014tja,Ita:2015tya,Larsen:2015ped}, the IBP relations in the Baikov representation are of type
\begin{align}
  0&=\int \d z_1 \cdots \d z_m \sum_{i=1}^m\frac{\partial}{\partial
     z_i} \bigg(a_i(z) P^\frac{D-L-E-1}{2} \frac{1}{z_1^{n_1} \cdots
     z_m^{n_m}}  \bigg),
\label{IBP_Baikov}
\end{align}
where each $a_i(z)$ is a polynomial in the variables $z_1,\dots, z_m$. Note that $P$ vanishes
on the boundary of the Baikov integration domain, so this form of IBP
identities does not have surface terms.

Suppose we wish to reduce an integral family with
$n_{j}\leq 0$, $j=\kappa+1,\ldots, m$, for some $\kappa$. That is, we face integrals with the inverse propagator product
$1/(D_1 \ldots D_\kappa)$ and the sub-topology integrals. We use the idea
of restricting to IBP systems without double propagators
\cite{Gluza:2010ws}, choosing suitable $a_i(z)$ to prevent the
appearance of double propagators in \eqref{IBP_Baikov}. In the Baikov
representation, we also need to avoid total derivatives with
dimension shifts \cite{Ita:2015tya,Larsen:2015ped}. These constraints
translate into syzygy equations of the following type: 
\begin{gather}
  \bigg(\sum_{i=1}^m a_i(z) \frac{\partial P}{\partial z_i} \bigg)+ b(z)
  P=0\,,
\label{module1}
\\
  a_i(z) = b_i(z) z_i \,,\quad i=1,\ldots, \kappa \,,
\label{module2}
\end{gather}
where $b(z)$ and the $b_i(z)$ are also polynomials in
$z_i$'s. Relation \eqref{module1} avoids dimension shifts of the integrals,
while \eqref{module2} ensures that there is no double
propagator for $D_i$ if the initial index $n_i=1$ in
\eqref{IBP_Baikov}. The goal is to find such
polynomials $a_i(z)$, $b(z)$, and $b_i(z)$. Since we require
polynomial solutions, this is not a linear algebra problem, but a
computational algebraic geometry problem.

We use the module intersection method from \cite{Zhang:2016kfo, Boehm:2018fpv} to solve \eqref{module1} and
\eqref{module2} simultaneously. Note that the analytic generators of
all solutions of
\eqref{module1} can be directly written down via either the canonical IBP
vector method \cite{Ita:2015tya} or the Gram matrix Laplace expansion
method \cite{Boehm:2017wjc}\footnote{We learned the Laplace expansion
  method from Roman Lee, and proved its completeness via the
  Gulliksen-Negard/Jozefiak exact sequence in \cite{Boehm:2017wjc}.}. The relations in
\eqref{module2}  can be trivially expressed as a module membership condition. Hence without any algorithmic computation, we know the individual
solutions for \eqref{module1} and \eqref{module2}, respectively. These form
polynomial submodules $M_1$ respectively $M_2$ of $R^m$ over the
polynomial ring $R=\mathbb Q(c_1,
\ldots, c_k)[z_1,\ldots, z_m]$ (where the variables $ c_1,\ldots c_k$ collect the Mandelstam
variables and the mass parameters). 
The task is then to compute
\begin{equation}
  M_1 \cap M_2\,.
\label{intersection}
\end{equation}
This module intersection can be obtained by computing a module Gr\"obner
basis in a particular ordering \cite{Boehm:2018fpv}. One decisive
strategy is the {\it localization} technique, which allows us to compute
$M_1\cap M_2$ over the polynomial ring $\tilde R =\mathbb Q[c_1,
\ldots ,c_k, z_1,\ldots, z_m]$. In this manner, we treat kinematic
variables in the same way as the Baikov variables. This greatly speeds up the
intersection computation for multi-scale problems, but results in a
redundant generating system. The latter can be trimmed further by
importing the result back to $R^m$ and removing redundant generators by
checking the leading
monomials. This is powered by \textsc{Singular}'s command {\bf
  simplify}. Once $M_1\cap M_2$ is obtained, we know all simultaneous solutions for
\eqref{module1} and \eqref{module2}, and can use \eqref{IBP_Baikov} to get
IBPs without double propagators.

We emphasize that, although \eqref{module1} and \eqref{module2} were
originally designed for IBPs without double propagators, the
solutions of \eqref{module1} and \eqref{module2} can be used to
simplify IBP systems {\it with} double or multiple propagators. Using
these solutions $a_i(z)$, the resulting IBP system does not
introduce integrals with higher powers of propagators, and hence also
greatly decreases the size of the IBP system.

Frequently, instead of computing IBPs directly, we compute IBPs on
spanning cuts and assemble the full IBPs afterwards. This amounts to
setting some of the $z_i$ to zero in \eqref{module1} and \eqref{module2}. For details on IBPs on cuts using the Baikov representation, we refer to~\cite{Boehm:2018fpv}.

Compared to the approach in \cite{Boehm:2018fpv}, we present the
following new features of the module intersection method in this
paper:
\begin{itemize}
\item When we compute the intersection $M_1\cap M_2$, instead of
  finding a full generating system, we heuristically impose a polynomial degree bound in
  the computation. Then we reduce the resulting IBPs over
  finite fields to test if we already have all the IBP relations
  needed. If the IBP relations are insufficient, we increase the
  degree bound and repeat the computation. This approach speeds up the
  intersection computation dramatically in many cases. In practice,
  we use the option {\bf degbound} in the computer algebra software
  {\sc Singular} to set the degree bound.
\item In the approach of \cite{Boehm:2018fpv}, the module intersection was only
  computed for the top sector, which, for the hexagon box diagram, turned out  to be sufficient for reducing
  integrals to a master integral basis. However, in this paper, we compute the module intersection for
  the top sector and also all subsectors. This approach may, in general, generate more IBP
  relations. Via linear algebra trimming as discussed in the next subsection,
  this approach eventually gives a block triangular linear system and
  makes the linear reduction easier.
\end{itemize}

\subsection{Linear reduction}
\label{linear_reduction}
For the simplified IBP system arising from the module-intersection method, we use
our own linear reduction algorithm to reduce the IBP system. The steps are:
\begin{enumerate}
\item Trim the linear system in two stages: (a) Set all the kinematic variables to integer
  values, and use linear algebra over a finite field to find the
  independent IBP relations. (b) Again over a finite field, carry out
  the reduction. From the intermediate steps, determine a
  sufficient subset of IBP relations for reducing the target
  integrals. These operations are powered by the finite field
  computation tool {\sc SpaSM}~\cite{spasm}.
\item Remove the overlap between two different cuts and simplify the
  linear system: If two cuts have a common master integral, use
  the idea from \cite{Chawdhry:2018awn} to set the master integral to zero in the
  IBP system of one of the two cuts. This will later on dramatically simplify the IBP
  reduction for the cut.
\item For the linear system simplified by the first two steps, we
  use our own {\sc Singular} row reduce echelon form (RREF) code over function fields to reduce the target
  integrals to master integrals. Our code applies both row and column
  swaps for finding the optimal pivots. Note that column swaps
  change the set of master integrals. After the RREF computation, we
  convert the new master integrals to the original master
  integrals. We have observed that this approach is in general much faster than
  fixing the column ordering and directly reducing the target
  integrals to the original master integrals.
\end{enumerate}

For difficult IBP reduction computations, we use a ``semi-numeric'' approach:
This approach sets several but usually not all of the kinematic
variables for the reduction computation to numeric values (that is, to constant integers). Without loss of generality, for the
kinematic variables $(c_1, \ldots, c_k)$, we set
\begin{equation}
  \label{eq:2}
     c_i \mapsto a_i, \quad 1\leq i \leq k_1,
\end{equation}
for some $k_1<k$ and some $a_i\in\mathbb Z$.

The actual degree of the coefficients in these variables can be decided by a univariate analytic computation (that is, we set all but one of the $c_i$ to constant values). For example, we may pick
the dimension $D$   and all parameters $c_i$ except $c_1$ as random integers, and then carry
out the reduction. This computation is much easier than the
actual IBP reduction with fully analytic parameters. From the
reduction, we determine the degree of $c_1$ in the final
IBP reduction coefficients. Proceeding similarly for each $i$, we find the degree of each $c_i$. This determines
the minimal number of semi-numeric points for the subsequent
interpolation step. (See \cite{Peraro:2016wsq} for an alternative way of
finding the degree of each parameter in a rational function.)

After accumulating enough
points, we collect the semi-numeric reduction
results and interpolate to get the final IBP reduction
coefficients. To do this, we first run step 3 above for a semi-numeric
set of parameters, find the optimal pivots and record the row/column swap history as
a \emph{trace} of our computation. For other numeric values, we always use
the same trace to 
ensure 
the relatively uniform running time of the computation. 

In practice, we use our rational function interpolation algorithm described
in Appendix~\ref{sec:appendix}. We do a reduction computation,  with a carefully chosen 
semi-numeric reference point,
\begin{equation}
  \label{eq:13}
  c_j \mapsto b_j , \quad b_j \in \mathbb Z, \quad k_1<j\leq k,
\end{equation}
and $c_1, \ldots c_{k_1}$ symbolic. Using the reference point result,  we convert the rational
function interpolation problem to individual polynomial interpolation
problems for the numerators
and denominators. With this approach, the number of
``semi-numeric'' computations is
\begin{equation}
  \label{eq:14}
  (d_1+1) \times (d_2+1) \times \ldots \times (d_{k_1}+1), 
\end{equation}
where the $d_i$, for $1\leq i \leq k_1$, are the maximal degrees of the
$c_i$ in the numerator and denominator polynomials in the RREF matrix. This algorithm is also implemented in  {\sc Singular}.

For the semi-numerical reduction and interpolation, we need
to parallelize our computations in an efficient way. Furthermore, with
semi-numeric points, we may have some bad points in the
reduction or interpolation. In order to make use of massively parallel computations in an efficient way, and to automize the workflow for the replacement of bad
points, we use the
modern workflow management system \textsc{GPI-Space}, in conjunction with the computer algebra system \textsc{Singular}. We will discuss the ideas behind this approach in the subsequent section.

\section{Massively parallel computations using \textsc{Singular} and GPI space} \label{sec:gpising}
~~

Large scale calculations such as row reductions of IBP identities in the case of Feynman diagrams which are relevant to current research in high-energy physics, are only feasible by using parallel computing on high-performance clusters. The computer algebra methods applied in this context require to model algorithms which rely on sub-computations with time and memory requirements that are difficult
to predict.  This is due, for example, to the behaviour of Buchberger's algorithm for finding Gr\"obner bases: Although this algorithm performs well in many practical examples of interest, its worst case complexity is doubly exponential in the number of variables~\cite{Mayr1982TheCO}. Nevertheless it turned out recently \cite{BDFPRR, mptopcom} that massively parallel methods, which have been a standard tool in numerical simulation for many years, can also be applied successfully in symbolic computation. Proposing the general use of massively parallel methods in computer algebra, we describe our ongoing effort in this direction which is based on connecting the computer algebra system \textsc{Singular}  for polynomial calculations with the workflow management \textsc{GPI-Space}. The latter consists of a scheduler distributing the actual computations to workers in the cluster, a virtual memory layer to facilitate communication between the workers, and a workflow management system which relies on modeling algorithms in terms of Petri nets. 

In its basic form, a Petri net is a directed bipartite graph with two kinds of nodes: While a \textit{place} can hold a number of indistinguishable (structure-less) \textit{tokens}, a \textit{transition} may \textit{fire} if each input place contains at least one token (we then say that the transition is \textit{enabled}). When fired, a transition  consumes one token from each input place and puts one token on each output place. See Figure \ref{fig enabled} for an enabled transition and its firing, and Figure \ref{fig disabled} for a transition which is not enabled. In the figures, places are shown as circles, transitions as rectangles, and tokens as black dots.

   \begin{figure}[h]
   \begin{center}
  \begin{tikzpicture}
  \node[transition] (1) {};
  \node[invisible]  (i) [right of=1] {};
  \node[invisible]  (ii) [left of=1] {};
  \node[place]      (2) [above of=i,node distance=0.5\nd] {};
  \node[place]      (3) [below of=i,node distance=0.5\nd] {};
  \node[place,tokens=1]      (4) [above of=ii,node distance=0.5\nd] {};
  \node[place,tokens=1]      (5) [below of=ii,node distance=0.5\nd] {};
  \path[->]
        (4) edge (1)
        (5) edge (1)
        (1) edge (2)
        (1) edge (3)
  ;
\end{tikzpicture}
\hspace{1cm}
\raisebox{7mm}{$\longmapsto$}
\hspace{1cm}
  \begin{tikzpicture}
  \node[transition] (1) {};
  \node[invisible]  (i) [right of=1] {};
  \node[invisible]  (ii) [left of=1] {};
  \node[place,tokens=1]      (2) [above of=i,node distance=0.5\nd] {};
  \node[place,tokens=1]      (3) [below of=i,node distance=0.5\nd] {};
  \node[place]      (4) [above of=ii,node distance=0.5\nd] {};
  \node[place]      (5) [below of=ii,node distance=0.5\nd] {};
  \path[->]
        (4) edge (1)
        (5) edge (1)
        (1) edge (2)
        (1) edge (3)
  ;
\end{tikzpicture}
\end{center}
\smallskip
  \caption{An enabled transition and its firing.\label{fig enabled}}
\end{figure}
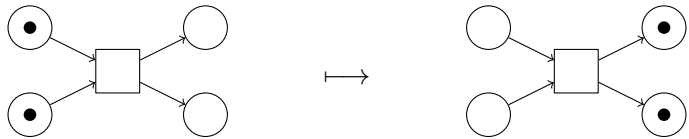

   \begin{figure}[h]
   \begin{center}
  \begin{tikzpicture}
  \node[transition] (1) {};
  \node[invisible]  (i) [right of=1] {};
  \node[invisible]  (ii) [left of=1] {};
  \node[place]      (2) [above of=i,node distance=0.5\nd] {};
  \node[place]      (3) [below of=i,node distance=0.5\nd] {};
  \node[place,tokens=1]      (4) [above of=ii,node distance=0.5\nd] {};
  \node[place]      (5) [below of=ii,node distance=0.5\nd] {};
  \path[->]
        (4) edge (1)
        (5) edge (1)
        (1) edge (2)
        (1) edge (3)
  ;
\end{tikzpicture}
\end{center}
\smallskip
  \caption{A transition which is not enabled.\label{fig disabled}}
\end{figure}
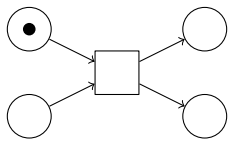
  
    The execution of a Petri net is non-deterministic: At each step, a single random enabled transition is chosen to fire. We have observed that the randomized reformulation of deterministic algorithms in computer algebra  in terms of Petri nets can lead to a more consistent and predictable behavior throughout the course of the computation. 

In our approach, we model the coarse-grained structure of an algorithm in terms of a Petri net. The transitions call procedures from the C-library version of \textsc{Singular} to do the actual computations. The result of this setup is a flexible framework for massively parallel computations in computational algebraic geometry (similar setups are possible using C-libraries of computer algebra systems aiming at possibly different application areas). Our framework has, for example, already been used to implement a non-singularity test for algebraic varieties \cite{Ristau,BDFPRR}, the computation of combinatorial objects in geometric invariant theory \cite{Reinbold}, and the computation of tropical varieties associated to algebraic varieties \cite{Bendle}. 
 
 For the efficient use in practical programming, the basic concept of a Petri net has to be extended. Here, \textsc{GPI-Space} provides multiple additional features:   
 \begin{itemize}
\item  Modeling complex algorithms just by the use of structure-less tokens is not very efficient. In \textsc{GPI-Space}, tokens can have a data type and hold actual data. In fact, it is often more efficient if the tokens just hold a reference to a storage place for the data (in memory or in the file system). Using the shared memory subsystem of \textsc{GPI-Space} or the powerful file systems of modern high-performance clusters, computations can then scale far beyond the limitations of a single machine.
 
\item The firing of a  transition may be subject to conditions which have to be fulfilled by the input tokens.
 
\item  Transitions in practice involve computations which take time. The properties of Petri nets allow us to execute different enabled transitions at the same time (\textit{task parallelism}) and to execute multiple instances of the same transition in parallel, provided the input places hold multiple tokens (\textit{data parallelism}). In Figure~\ref{fig par}, the transitions $f_1$ and $f_2$ can fire in parallel, and, if the input place of $f_i$ holds multiple tokens, then $f_i$ can fire in multiple instances.
  \begin{figure}[h]
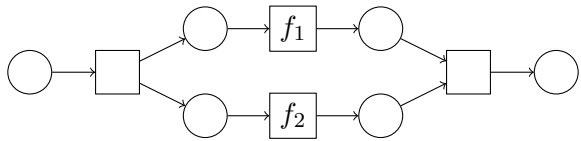

  \begin{petrinet}
  \node[place]      (0)              {};
  \node[transition] (1) [right of=0] {};
  \node[invisible]  (i) [right of=1] {};
  \node[place]      (2) [above of=i,node distance=0.5\nd] {};
  \node[place]      (3) [below of=i,node distance=0.5\nd] {};
  \node[transition] (f) [right of=2] {$f_1$};
  \node[transition] (g) [right of=3] {$f_2$};
  \node[place]      (4) [right of=f] {};
  \node[place]      (5) [right of=g] {};
  \node[transition] (j) [right of=i,node distance=3\nd] {};
  \node[place]      (6) [right of=j] {};
  \path[->]
        (0) edge (1)
        (1) edge (2)
        (1) edge (3)
        (2) edge (f)
        (f) edge (4)
        (3) edge (g)
        (g) edge (5)
        (4) edge (j)
        (5) edge (j)
        (j) edge (6)  
  ;
\end{petrinet}
\smallskip
  \caption{Task and data parallelism in a Petri net.\label{fig par}}
\end{figure}
\end{itemize}

We have observed that some algorithms in computer algebra scale in a
superlinear way when implemented in parallel as a Petri net. The
reason is that then, at run time,  the algorithms can automatically
determine from a given set of paths a path which leads to the solution in the fastest possible
way (see \cite[Section 6.2]{BDFPRR}).
 
 In the next section, we illustrate the use of the \textsc{Singular}-\textsc{GPI-Space} framework for applications in high-energy physics by modeling our IBP reduction algorithm.

\section{Parallel matrix reduction as a Petri net}\label{sec IBP Petri} 

In this section, we first describe how to model the parallel IBP reduction algorithm in terms of a Petri net. Focusing on the cut $\{1,3,4,5\}$ of the two-loop five-point nonplanar double pentagon diagram, we then discuss timings and scaling of the algorithm to indicate the practical use and significant potential of the \textsc{Singular}-\textsc{GPI-Space} framework for algorithmic problems in high-energy physics.

\subsection{General structure of the algorithm}
Our approach includes a massively parallel execution of row-reductions over function fields, where a number of parameters has been replaced by integers, followed by a parallel interpolation step to reconstruct the dependency on these parameters.

So the task is to find the reduced row-echelon form $M_\text{red}$ of a large linear system of equations, given as a matrix $M$ over the rational function field $\mathbb Q(c_1, \dots, c_k)$. Since applying Gaussian elimination directly is not feasible, we instead proceed by substituting, say, the first $r$ parameters by the coordinates of a point $a \in \mathbb Z^r$, and then by computing the reduction
\[
  (M|_{c_1\mapsto a_1, \dots, c_r\mapsto a_r})_{\text{red}}.
\]
We refer to Section~\ref{linear_reduction} above for details on how we handle this reduction step. To determine the number of interpolation points required to reconstruct the dependency on $c_1, \dots, c_r$, we find bounds for the degrees of numerators and denominators for each parameter by doing a univariate row reduction (that is, all but one of the parameters are set to be numeric). After the reduction, we check that the resulting matrix is equal to the desired result
\[
  {M_{\text{red}}}|_{c_1\mapsto a_1, \dots, c_r\mapsto a_r}
\]
by normalizing it relative to a previously computed reference matrix with $c_{r+1}, \dots, c_k$ constant, and performing degree checks using the exact degrees obtained from the univariate calculations. These steps are described in more detail in Appendix~\ref{sec:rational_func_interpolation}. The final result $M_\text{red}$  is then found by iteratively combining the reduced matrices via univariate interpolation (see again Appendix~\ref{sec:rational_func_interpolation}).

Let $d_1, \dots, d_r$ be degree bounds for the entries of $M_{\text{red}}$ in the parameters $c_1, \dots, c_r$, respectively. To obtain $M_{\text{red}}$ by interpolation, we need $d_1+1$ matrices over $\mathbb Q(c_2, \dots, c_k)$ of the form
\begin{equation}
  {M_{\mathrm{red}}}|_{c_1\mapsto a_1^{(0)}}, \dots,
  {M_{\mathrm{red}}}|_{c_1\mapsto a_1^{(d_1)}},
\end{equation}
for $d_1+1$ values $a_1^{(0)}, \dots, a_1^{(d_1)} \in \mathbb Z$. Similarly, to obtain any one of the above matrices, we need $d_2+1$ matrices over $\mathbb Q(c_3, \dots, c_k)$. Continuing inductively, this process ends with matrices defined over $\mathbb Q(c_{r+1}, \dots, c_k)$, which are then computed by reduction with $c_1, \dots, c_r$ numeric. This tree-like dependency structure is depicted in Figure~\ref{fig:inttree}.

\subsection{Managing the interpolation}\label{sec:intstor}

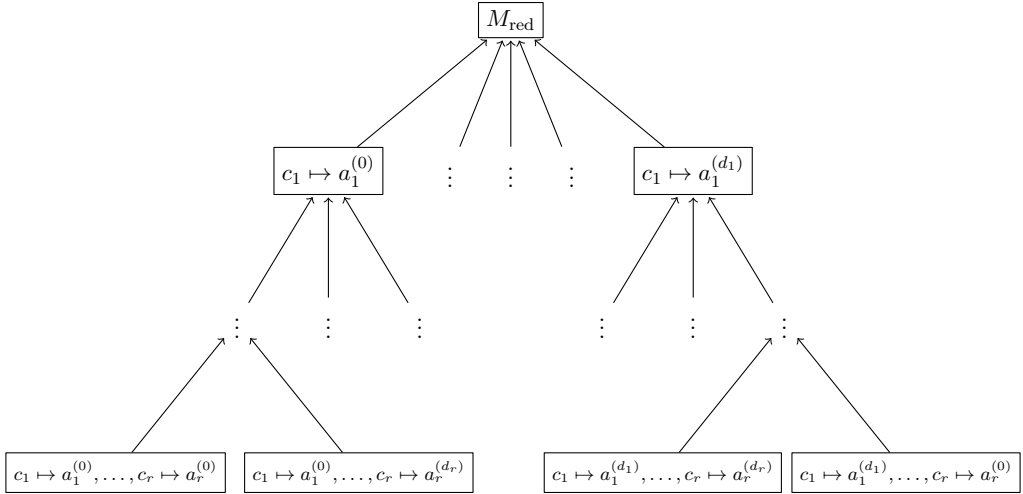
\begin{figure}
  \centering
  \begin{tikzpicture}[scale=0.8, every node/.style={scale=0.8}]
    \node[transition] at (0,0) (root) {$M_{\text{red}}$};
    \node[transition] at (-3,-2.5) (n11) {${c_1 \mapsto a_1^{(0)}}$}
      edge [post] (root);
    \node[transition] at (3,-2.5) (n1d) {${c_1 \mapsto a_1^{(d_1)}}$}
      edge [post] (root);
    \node at (-1,-2.5) {$\vdots$} edge [post] (root);
    \node at (0,-2.5) {$\vdots$} edge [post] (root);
    \node at (1,-2.5) {$\vdots$} edge [post] (root);

    \node at (-4.5,-5) (nr1) {$\vdots$} edge [post] (n11);
    \node at (-3,-5) {$\vdots$} edge [post] (n11);
    \node at (-1.5,-5) {$\vdots$} edge [post] (n11);

    \node at (4.5,-5) (nrd) {$\vdots$} edge [post] (n1d);
    \node at (3,-5) {$\vdots$} edge [post] (n1d);
    \node at (1.5,-5) {$\vdots$} edge [post] (n1d);

    \node[transition] at (-6.5,-7.5) (nd1-1)
      {\footnotesize${c_1 \mapsto a_1^{(0)}, \dots, c_r \mapsto a_r^{(0)}}$}
      edge [post] (nr1);
    \node[transition] at (-2.5,-7.5) (ndd-1)
      {\footnotesize${c_1 \mapsto a_1^{(0)}, \dots, c_r \mapsto a_r^{(d_r)}}$}
      edge [post] (nr1);

    \node[transition] at (6.5,-7.5) (nd1-d)
      {\footnotesize${c_1 \mapsto a_1^{(d_1)}, \dots, c_r \mapsto a_r^{(0)}}$}
      edge [post] (nrd);
    \node[transition] at (2.5,-7.5) (ndd-d)
      {\footnotesize${c_1 \mapsto a_1^{(d_1)}, \dots, c_r \mapsto a_r^{(d_r)}}$}
      edge [post] (nrd);
  \end{tikzpicture}
  \caption{The structure of the interpolation tree.}
  \label{fig:inttree}
\end{figure}

We model the current status of the interpolation process in a tree-like data structure corresponding to that from Figure~\ref{fig:inttree}, with references to the reduction results  at the leaves, and  references to the interpolation results at the other nodes. Within \textsc{GPI-Space}, reductions and interpolations are executed according to this data structure. The tree is generated as soon as the degree bounds $d_1, \dots, d_r$ are known, and it is extended if the algorithm requires additional data points due to the occurrence of bad interpolation points.

\subsection{Description of the Petri net} \label{sec:algonet}

Figure~\ref{fig:fullnet} depicts the Petri net that implements the complete reduction algorithm. Going beyond the standard syntax introduced in Section~\ref{sec:gpising}, dashed arrows stand for read-only access, that is, the data in the respective places is not consumed. The dotted arrows illustrate read and write access to the interpolation tree described in Section \ref{sec:intstor}. A transition can be annotated by conditions which indicate that the transition can only fire by consuming tokens for which the conditions evaluate to true.\footnote{When formulating conditions
in the Petri net, we use the name of a place and a token on the place interchangeably.}  In the following, we describe the individual structures of the net:

\begin{figure}[tbhp]
  \centering
  \begin{tikzpicture}
    \tikzset{>=latex}
    \tikzset{transition/.append style={font=\ttfamily}}
    \node[place] at (0,1.5) (indata) {$I$};
    \node[transition,align=center] at (-3.0,0) (trace)
      {trace\\{\scriptsize if not $I$.trace exists}}
      edge [pre,out=90,in=180] (indata);
    \node[transition,align=center] at (3.0,0) (copy)
      {copy\\{\scriptsize if $I$.trace exists}}
      edge [pre,out=90,in=0] (indata);
    \node[place] at (0,0) (tempdata) {}
      edge [pre] (copy)
      edge [pre] (trace);

    \node[transition] at (0,-1.5) (init) {init}
      edge [pre] (tempdata);

    \node[place] at (0,-3) (data) {$I_t$} edge [pre] (init);
    \node[place] at (-4,-1.5) (init1) {}
      edge[pre] (init);
    \node[place] at ( 4,-1.5) (init2) {}
      edge[pre] (init);

    \node[transition] at (-4, -3) (cdeg) {degrees}
      edge [out=90,in=-90,pre] (init1)
      edge [pre,dashed] (data);
    \node[transition] at (4, -3) (cref) {reference}
      edge [out=90,in=-90,pre] (init2)
      edge [pre,dashed] (data);

    \node[place] at (4,-4.5) (ref) {}
      edge [pre] (cref);
    \node[transition,align=center] at (4,-6) (normalize)
      {normalize\\{\scriptsize if $m$.valid}}
      edge [pre,dashed] (ref);

    \node[place] at (-4, -4.5) (vdeg) {$d_v$}
      edge [pre] (cdeg);
    \node[place] at (0, -4.5) (mdeg) {$d_m$}
      edge [pre,in=-40,out=180] (cdeg)
      edge [out=0,in=140,post,dashed] (normalize);

    \node[database,align=center] at (-4,-8.5) (stor) {interpolation tree};

    \node[transition] at (-4, -6) (gpts) {points}
      edge [stealth-stealth,dotted,thick] (stor)
      edge [pre] (vdeg);
    \node[place] at (-1.5, -6) (p) {$p$}
      edge [pre] (gpts);
    \node[transition] at (0, -6) (red) {reduce}
      edge [pre] (p)
      edge [out=165,in=-165,color=white,line width=4pt,shorten <=5pt] (data)
      edge [out=165,in=-165,pre,dashed] (data);
    \node[place] at (1.5,-6) (redm) {$m$}
      edge [pre] (red)
      edge [post] (normalize);
    \node[transition,align=center] at (1.5,-7.5)
      {replace failure\\{\scriptsize if not $m$.valid}}
      edge [pre] (redm)
      edge [stealth-stealth,out=182,in=10,dotted,thick] (stor)
      edge [out=178,in=-60,post] (p);

    \node[place] at (4,-8.75) (norm) {$n$}
      edge [pre] (normalize);
    \node[transition,align=center] at (1.5,-8.75)
      {replace invalid\\{\scriptsize if not $n$.valid}}
      edge [pre] (norm)
      edge [out=178,in=-120,color=white,line width=4pt,shorten <=5pt] (p)
      edge [out=178,in=-120,post] (p)
      edge [stealth-stealth,out=182,in=0,dotted,thick] (stor);

    \node[transition,align=center] at (0,-10.25) (storN)
      {store normalized\\{\scriptsize if $n$.valid}}
      edge [stealth-stealth,out=180,in=-45,dotted,thick] (stor)
      edge [out=0,in=-90,pre] (norm);
    \node[place] at (0,-12) (ipt) {$i$}
      edge [pre] (storN);
    \node[transition,align=center] at (2.5,-12)
      {discard\\{\scriptsize if not $i$.valid}}
      edge [pre] (ipt);
    \node[transition,align=center] at (-2.5,-12) (ip)
      {interpolate\\{\scriptsize if $i$.valid}}
      edge [stealth-stealth,out=90,in=-66,dotted,thick] (stor)
      edge [pre] (ipt);

    \node[transition] at (-1.95,-13.5) (storI) {store interpolated}
      edge [stealth-stealth,out=165,in=-90,dotted,thick] (stor)
      edge [out=10,in=-90,post] (ipt);
    \node[place] at (-4.75,-12.75) {}
      edge [in=180,out=66,color=white,line width=4pt] (ip)
      edge [in=180,out=66,pre] (ip)
      edge [in=180,out=-66,post] (storI);
  \end{tikzpicture}
  \caption{The Petri net for row reduction via interpolation. A description of the syntax is given in the first paragraph of Section~\ref{sec:algonet}.}
  \label{fig:fullnet}
\end{figure}
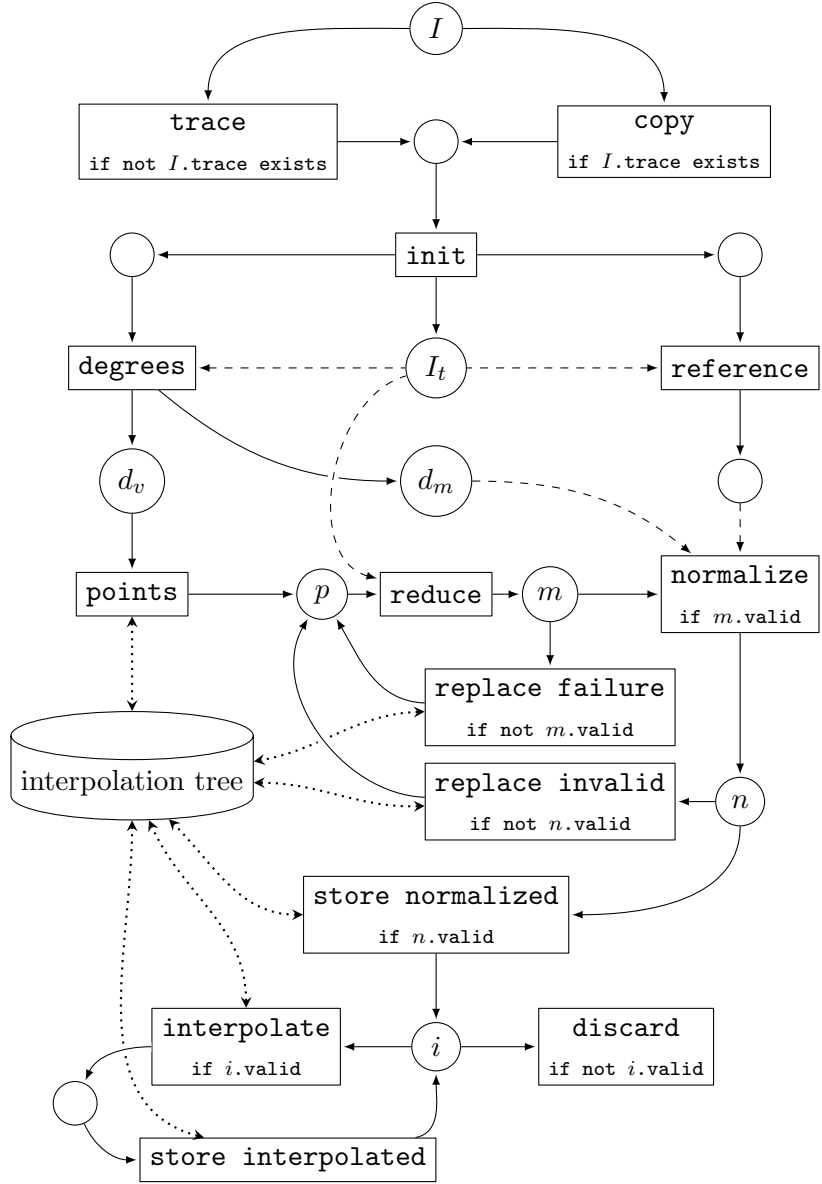

\begin{description}[leftmargin=0cm]
  \item[Input token:] The net is initialized with one token: A token on the place $I$, which holds references to the following input data:
    \begin{itemize}
      \item The input linear relations, which are given as a matrix $M$ over the rational function field $\mathbb Q(c_1, \dots, c_k)$.
      \item The vector of indices of the parameters which will be interpolated (in the following we assume that these indices are $1, \dots, r$).
      \item The vector of indices of the target variables.
      \item Optionally: A precomputed trace for the reduction step (consistent with the targets). In the Petri net, the trace is referred to as \texttt{$I$.trace} (we use the usual dot-notation for sub-data structures). Note that the trace fixes the variables corresponding to the master integrals.
    \end{itemize}
  \item[Transition \texttt{trace}:] If the token on $I$ does not contain a trace, then \texttt{trace} is enabled, computes a trace for the linear reduction (see Section \ref{linear_reduction}) and returns a copy of $I$ with the trace included.
  \item[Transition \texttt{copy}:] If the token on $I$ already contains a trace, then \texttt{copy} is enabled and simply passes the token on $I$ through.
  \item[Transition \texttt{init}:] This transition takes the input token, which was produced by either \texttt{trace} or \texttt{copy}, and pushes it onto $I_t$. This way, the input data on $I_t$ is guaranteed to contain trace data. It additionally enables the transitions \texttt{degrees} and \texttt{reference}.
  \item[Transition \texttt{reference}:] This generates a random substitution point $q = (q_{r+1}, \dots, q_{k})$ with values for all parameters which will \emph{not} be interpolated, substitutes the $q_{i}$ for the $c_i$, and runs the row reduction step (see Section~\ref{linear_reduction}), that is, computes
    \[
      M_{\text{ref}} := {(M|_{c_{r+1}\mapsto q_{r+1}, \dots, c_k\mapsto q_k})}_{\text{red}}.
    \]
    The transition then stores the actual result in the file-system and produces an output token which contains both a reference to the result and the point $q$. The stored data will be used later in the normalization step of the interpolation (see above).
  \item[Transition \texttt{degree}:] This generates a substitution point $p^{(j)} \in \mathbb Z^{\{1, \dots, j-1, j+1, \dots, k\}}$ for each $1\leq j \leq k$ yielding a matrix
    \[
      M^{(j)} := M|_{c_1\mapsto p^{(j)}_1, \dots, c_{j-1}\mapsto p^{(j)}_{j-1}, c_{j+1}\mapsto p^{(j)}_{j+1}, \dots, c_k \mapsto p^{(j)}_k}
    \]
    over the field $\mathbb Q(c_j)$. After applying the row reduction, $M^{(j)}_{\text{red}}$ can be used to determine degree bounds for the numerator and denominator of each entry of the final result $M_{\text{red}}$ as a polynomial in $c_j$.

    For $j \leq r$, we need a global degree bound to determine the number of interpolation points. We thus take the maximum of all numerator and denominator degrees of entries of $M^{(j)}_{\text{red}}$, and store these as a vector in $\mathbb N_0^{\{1, \dots, r\}}$, which is put on the place $d_v$.

    If $j > r$, two integer matrices will be produced, which store the degrees of the numerators and denominators of each
    entry of $M_{\text{red}}$, respectively. This information will be used later to filter out bad interpolation points, that is, points at which polynomial cancellation occurs (see Appendix~\ref{sec:rational_func_interpolation}). The result is stored in the file system and a token with a reference to the result is put on the place $d_m$.

    Note that \texttt{degree} is in fact modeled by a sub-Petri net which behaves in a hierarchical manner as a transition.
        In practice, we actually compute multiple matrices $M^{(j)}$ per parameter to reduce the probability of a bad point producing wrong degree bounds.
  \item[Transition \texttt{points}:] This transition takes the degree data in $d_v$ and initializes the interpolation tree described in Section~\ref{sec:intstor} and depicted in Figure~\ref{fig:inttree}. This, in turn, produces the corresponding set of interpolation points, which are put as separate tokens on the place $p$.
  \item[Transition \texttt{reduce}:] This transition consumes a point $p' \in \mathbb Z^{\{1, \dots, r\}}$ from the place $p$ and computes
    \[
      {(M_{c_1\mapsto p'_1, \dots, c_m \mapsto p'_r})}|_{\text{red}}.
    \]
    The resulting matrix together with its interpolation point are put on the place $m$.  Since \texttt{reduce} performs parameter substitutions in rational function expressions, the computation may fail due to division by zero. If this happens, \texttt{$m$.valid} is set to \texttt{false}, otherwise  it is set to \texttt{true}.
  \item[Transition \texttt{replace failure}:] An input token for which \texttt{$m$.valid} is \texttt{false}  is consumed by the transition \texttt{replace failure}, which marks the respective interpolation point as failed in the interpolation tree. If necessary, the interpolation tree is extended by additional interpolation points,  which are also put on the place $p$.
  \item[Transition \texttt{normalize}:] An input token for which \texttt{$m$.valid} is true  is consumed by the transition \texttt{normalize}. This transition reads $M_{\text{ref}}$ and multiplies the input matrix referenced by $m$ with a suitable constant factor. It also compares the entries with the degree matrices in $d_m$ to identify bad interpolation points. The result is put on the place $n$. If the corresponding point was bad, n.\texttt{valid} is set to \texttt{false}, otherwise to \texttt{true}.
  \item[Transition \texttt{replace invalid}:] For an input token for which  n.\texttt{valid} is \texttt{false}, the transition generates new interpolation points in a fashion similar to that in \texttt{replace failure}.
  \item[Transition \texttt{store normalized}:] For an input token for which  n.\texttt{valid} is \texttt{true}, the transition marks the corresponding interpolation point as successful in the external storage. If enough interpolation points for a given parameter have been marked as successful, the storage produces a token on place $i$, which triggers the respective interpolation. If the point $(p'_1, \dots, p'_r)$ triggers the interpolation (which will then use further points of the form $(p'_1, \dots, p'_{r-1},p''_r)$), the result of the interpolation will be associated to the point $(p'_1, \dots, p'_{r-1})$ in the interpolation tree. If there are not yet enough interpolation points, the transition produces a token which only contains \texttt{$i$.valid} with value \texttt{false}. 
  \item[Transition \texttt{discard}:] This transition discards tokens with \texttt{$i$.valid} equal to  \texttt{false}.
  \item[Transition \texttt{interpolate}:] Tokens with \texttt{$i$.valid} equal to  \texttt{true} are consumed by this transition, which then retrieves the references to the input data for the interpolation from the interpolation tree, loads the respective data from the file system, and executes the interpolation. If (in the above notation) the token holds $(p'_1, \dots, p'_{r-1})$, then for $(d_v)_{r}+1$ many points the corresponding row reduced matrices are retrieved from the storage. Note that due to the tree structure of the interpolation tree, all these points must have the first $r-1$ coordinates equal to $(p'_1, \dots, p'_{r-1})$. The interpolation is then performed entry-wise as explained in Appendix~\ref{sec:rational_func_interpolation}.
  \item[Transition \texttt{store interpolated}:] This transition marks the current point $(p'_1, \dots, p'_{r-1})$ in the interpolation tree as processed. If $r>1$, just like in \texttt{store normalized}, the transition  produces an interpolation token for the next parameter. If $r=1$, we have arrived at the final result, and a token with \texttt{$i$.valid} equal to  \texttt{false} is produced, which will then be discarded.

  The Petri net contains additional infrastructure (not described here) which terminates the execution once  no tokens exist any more on the places $i$ and $p$.
\end{description}

\subsection{Parallel timings} \label{sec:timing}

To illustrate the efficiency of our approach, we consider the cut $\{1,3,4,5\}$ of the double pentagon diagram  (see Section~\ref{sec:doublepentagon} for a discussion of all possible cuts). Choosing this particular cut, which is less complex than others, our computations finish even when only a small number of cores is involed.
This is necessary to analyze the scaling of our algorithm. In Table~\ref{tab:timings1345}, we give timings for different numbers of cores. All timings are in seconds, taken on the high performance compute cluster at the Fraunhofer Institute for Industrial Mathematics (ITWM). Each compute node provides two Xeon E5-2670 processors, which amounts to $16$ cores\footnote{Hyperthreading is disabled.} running at a base clock speed of 2.6\,GHz. Each node has 64\,GB of memory. For all runs with more than 15 cores, on each node we ran 15 compute jobs and one job for interfacing with the storage system. Since the storage jobs use negligible computation time, we omit them from the CPU core count when determining speedup and efficiency.

\begin{table}[tbhp]
  \centering
  \begin{tabular}{r|r||r|r|r}
  &&& \multicolumn{2}{c}{\textbf{relative}} \\
    \textbf{nodes} & \textbf{cores} & \textbf{runtime} & \textbf{speedup} & \textbf{efficiency} \\
    \hline
    1 & 1 & 122857.6 & 1.201 & 1.201 \\
    1 & 15 & 9837.8 & 15.000 & 1.000 \\
    2 & 30 & 4954.8 & 29.7822 & 0.992 \\
    4 & 60 & 2625.4 & 56.2058 & 0.936 \\
    8 & 120 & 1341.3 & 110.014 & 0.916 \\
    14 & 210 & 952.3 & 154.958 & 0.737 \\
    15 & 225 & 705.6 & 209.132 & 0.929 \\
    16 & 240 & 694.3 & 212.514 & 0.885 \\
    29 & 435 & 611.8 & 241.199 & 0.554 \\
    30 & 450 & 385.4 & 382.856 & 0.850 \\
    32 & 480 & 379.9 & 388.336 & 0.809 \\
    40 & 600 & 367.7 & 401.307 & 0.668 \\
    48 & 720 & 363.2 & 406.195 & 0.564 \\
  \end{tabular}
  \caption{Timings and efficiency for the cut $\{1,3,4,5\}$. We use the same algorithm for all core counts. The single core run serves as a reference.}
  \label{tab:timings1345}
\end{table}

Apart from the running time $T(n)$ of the algorithm on a total of $n$ cores, we also give the speedup $S(n) = \smash{\frac{T(1)}{T(n)}}$ and the efficiency $E(n) = \smash{\frac{T(1)}{nT(n)}}$, which measure how \enquote{well} the algorithm parallelizes with increasing core counts. Note that the single-core timing is somewhat special: As experiments have shown, the performance per core decreases with the number of cores used on a given node. This effect has been investigated in \cite{BDFPRR} (see in particular \cite[Figure~5]{BDFPRR}). Thus, for the analysis of the expected run-time below, we rather consider the \emph{relative} speedup and efficiency with respect to the 15-core timing. This in particular makes the assumption that the 15-core speedup is 15.

\begin{figure}[htbp]
  \begin{center}
    \input{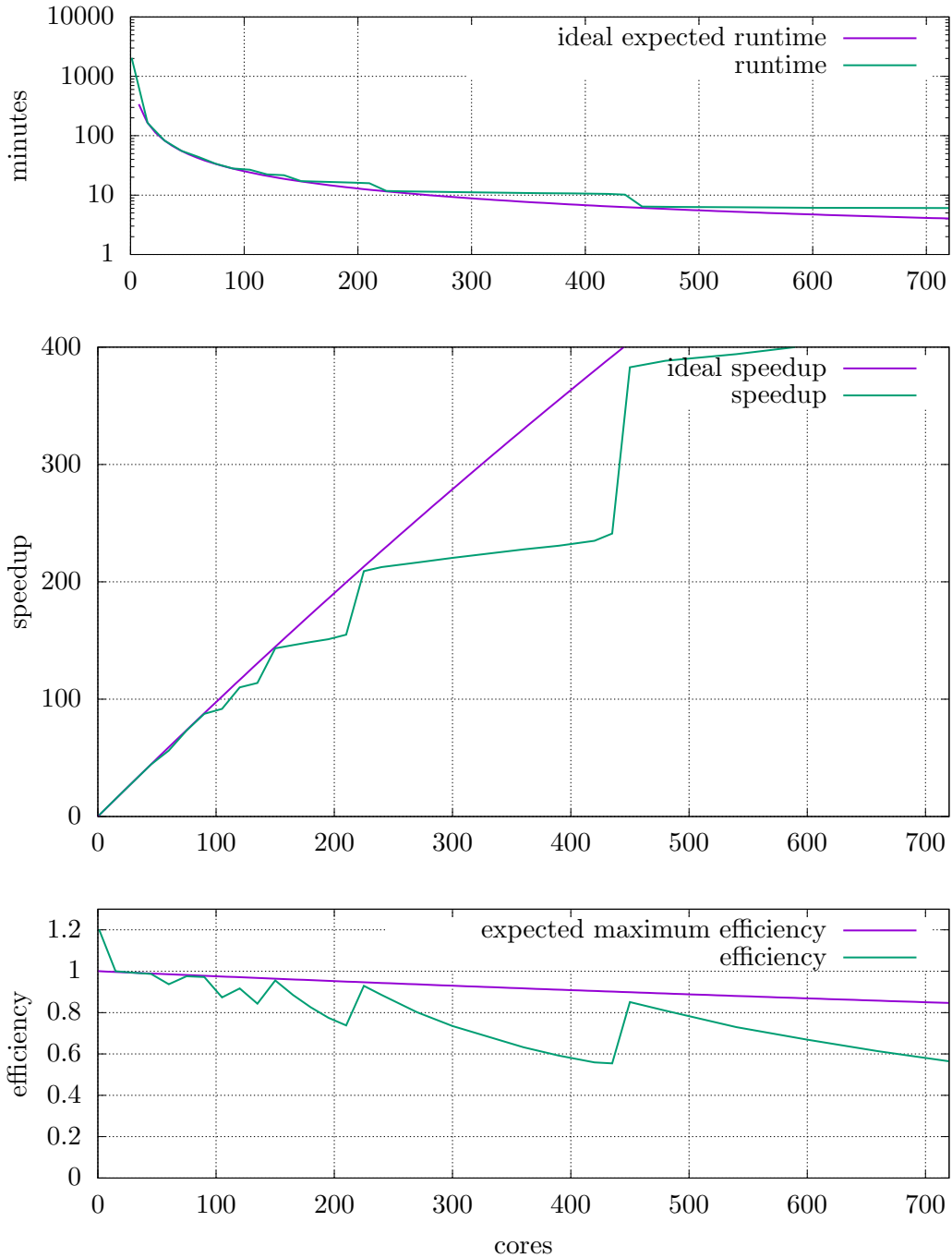}
    \vspace{-6mm}
  \end{center}
  \caption{Running time, relative speedup and efficiency graphs for cut $\{1,3,4,5\}$}
  \label{fig:time1345}
\end{figure}

The saw-tooth shape of the efficiency graph in Figure \ref{fig:time1345} (and the corresponding behavior in the timing and speedup graphs) is due to the fact that the number of reductions to execute is usually not divisible by the number of cores utilized. Since in our test problem approximately $450$ reductions are required to enable the final interpolation, the running time of the full algorithm is roughly
\begin{equation}
  \left\lceil\frac{450}{\text{number of CPUs}}\right\rceil.
\end{equation}
This effect can be avoided by a more fine-grained structuring of the problem (for instance by interpolating more parameters). Note, however, that increasing the number of processes in this way will lead to more overhead via inter-process communication and disk accesses. Thus, dividing the algorithm into very small parts may in fact slow down the overall computation.

Figure~\ref{fig:time1345} also depicts the ideal expected runtime, speedup and efficiency. These ideal graphs stem from the simple assumption, called \emph{Amdahl's law}, that an algorithm can be divided up into a part that is ideally parallelizable and a part which is not parallelizable at all. Denoting the parallelizable fraction by $f$, the expected runtime $T_{\text{ideal}}(n)$ on $n$ cores is not $\frac{T(1)}n$, but rather
\begin{equation}
  T_{\text{ideal}}(n) = (1-f)\cdot T(1) + \frac{f\cdot T(1)}n,
\end{equation}
which yields the ideal speedup and efficiency
\begin{equation}
  S_{\text{ideal}}(n) = \frac{n}{(1-f)\cdot n + f},
  \quad
  E_{\text{ideal}}(n) = \frac{1}{(1-f)\cdot n + f}.
\end{equation}
Using the experimental values for 15 and 30 cores, we arrive at a value $f \approx 0.999748$, that is, only 0.025\,\% of the algorithm is not parallelizable.

As we can see, the ideal curves give a fairly tight bound on the actual timings, at least in the cases where the core count is properly aligned to the number of reductions. This indicates that our approach for parallelization not only provides an automatic and fast solution to a tedious and complicated task, but stays highly efficient even when used with a large amount of computing power.

\section{IBP conversion between different integral bases}
\label{sec:IBP_conv} 
It is well known that the IBP coefficient size may vary significantly if we choose
different master integral bases. 
%
We prefer the IBP reduction to a uniformly transcendental (UT) basis as
introduced in \cite{Henn:2013pwa}, for several reasons: a) The differential equations satisfied by a UT
integral basis have a particularly simple form
\cite{Henn:2013pwa} which allows for the integrals to be solved
analytically in terms of polylogarithms. There is also evidence that
for numerical computations, a UT basis is more convenient to
evaluate\footnote{Private communication with Yang Gang.}. So the IBP
reduction to a UT basis greatly simplifies the amplitude computations
after the IBP reduction. b) We observe that, in the case of the double pentagon, the IBP
coefficients in a UT basis are significantly simpler than those in a
traditional Laporta basis. This makes the
IBP relations easier to use.

In practice, we consider special forms of UT bases, the so-called
dlog bases, which will be introduced in the next subsection.

\subsection{Dlog bases  and the dlog algorithm}
We say that a Feynman integral is a \emph{dlog integral} if its integrand 
\begin{equation}
  dI=dx_1 \wedge dx_2 \wedge ... \wedge dx_n R(x_1,...,x_n)\,,
\end{equation}
with $R(x_1,...,x_n)$  a rational function in $x_1, ..., x_n$,
can be expressed in dlog form \cite{Bern:2014kca}, that is, it can be written 
as a linear combination of type
\begin{equation}
\label{eq:dlogform}
  dI=\sum_i c_i\, d\textrm{log} f_{i,1} \wedge ... \wedge d\textrm{log} f_{i,n},
\end{equation}
with rational functions $f_{i,j}$ in $x_1, ..., x_n$. This is only possible if the integrand
has at most simple poles, including points at infinity.  For example, both forms $\frac{dx}{x^2}$ and 
$dx$ admit no dlog form because of the double poles at zero respectively infinity.

The coefficients $c_i$ in equation \eqref{eq:dlogform} are called leading
singularities \cite{ArkaniHamed:2010gh}. For Feynman integrals, that are not of the elliptic type, they are in general algebraic functions of the external variables. By choosing an appropriate parametrization of the external variables, the leading singularities are typically rational functions. This is, in particular, true for the two-loop five-point integrals that are discussed in the next section. The leading singularities can 
also be understood as integrals over the original integrand where the integration contour
is localized around the poles of the integrand. Leading singularities and the integrals
integrated on the real contour have analytic properties in common.
So, integrals with leading singularities that are just constant numbers are
particularly useful, most importantly because they fulfill
differential equations in the canonical form \cite{Henn:2013pwa}.
This implies that they have the property of
uniform transcendental weight, which means that if the series is expanded in $\epsilon$, the
parameter of dimensional regularization, the coefficients have homogeneous transcendental
weight and the weight increases by one for each order in~$\epsilon$.

Next, we recall from \cite{WasserMSc}  how to transform a given integrand
into dlog form, in case this is possible.
Given an integrand in $n$ integration variables, we choose, if possible, one variable $x$ that is linear
in all denominator factors and do a partial fraction decomposition while treating all other
variables as constants. In this way, we obtain a sum of integrands of the form
\begin{equation}
 \frac{dx}{x-a} \wedge \Omega = d\log(x-a)\wedge \Omega,
\end{equation}
where $\Omega$ is an $(n-1)$-form, independent of $x$, and $a$ is a polynomial
that may depend on the other integration variables.
Then we iterate this procedure taking $\Omega$ as our new integrand until
no integration variables are left. If in any intermediate step a pole of degree two or
higher is encountered, then the integrand does not admit a dlog form.
There are cases where no variable exists that is linear in all denominator factors. One
way to proceed in such a case is to make a variable transformation such that
at least one of the new variables is linear in all denominator factors.

The algorithmic approach of this section
was used in \cite{Chicherin:2018mue} and \cite{Chicherin:2018old}
to construct a complete basis of dlog master integrals
with constant leading singularities for all two-loop five point integral families.
The denominator structure for each integral family is given by the propagators.
To construct the dlog integrals
we make a general numerator ansatz. We write the numerator as a linear combination of terms that are products of inverse propagators
and irreducible scalar products. Each term is multiplied by a free parameter, and by applying
the algorithm to this general integrand, we can determine values of the free parameters
such that the integrand has a dlog form and constant leading singularities. In this way, we obtain a set of dlog integrals that form a basis of dlog master integrals.

In general, the dlog algorithm can be applied only in a dimension that is an integer number,
which we choose to be four. The loop momenta are very conveniently parametrized using
spinor helicity variables as in \cite{Bern:2014kca}. Although this parametrization can be very useful, it also has its limitations as soon as the numerator has terms that vanish in dimension four,
but which are non-zero in generic dimension $D$. In such cases, an extended approach as in \cite{Chicherin:2018old} using the Baikov parametrization  can be applied.

\subsection{IBP reduction with a dlog basis}
Given a dlog basis, we discuss the IBP reduction in two settings:
\begin{enumerate}
\item When both the IBP coefficients in the Laporta basis and the dlog
  basis are needed, we first compute the reduction in the Laporta
  basis $I$ with our module intersection and \textsc{GPI-Space} reduction
  algorithm,
  \begin{equation}
    \label{eq:1a}
    F=AI,
  \end{equation}
where $F$ is the list of target integrals as a column vector. Then we
reduce the dlog basis $\tilde I$ to the Laporta basis $I$,
 \begin{equation}
    \label{eq:1b}
    \tilde I=T I.
  \end{equation}
Note that since the dlog basis construction has a restriction on the numerator degree, this reduction is usually easy. Terms exceeding the allowed numerator degree have double poles at infinity. This can be seen by inverting the loop momenta $k_i^{\mu} \rightarrow k_i^{\mu}/k_i^2$. Using our
{\sc Singular} RREF code, with a good pivot strategy, we can analytically find the inverse
$T^{-1}$. The matrix product $A T^{-1}$ contains the coefficients
of an IBP reduction to the dlog basis.

We remark that the product $A T^{-1}$ can be difficult to calculate even if
$T^{-1}$ has a relative small size. Instead of computing the
product directly, we again use the semi-numerical approach, 
setting several of the kinematic values to be integers, computing the product
several times, and then using our interpolation program to get the fully
analytical matrix product $A T^{-1}$. This is again implemented
using our \textsc{Singular}-\textsc{GPI-Space} framework.  

\item When only the IBP coefficients in a dlog
  basis are needed, we apply our semi-numerical reduction method to
  a set of numeric IBP coefficients in the Laporta basis. Instead of
  interpolating these coefficients, we use the semi-numeric points to interpolate the product $A T^{-1}$, not calculating the analytic form of $A$.
\end{enumerate}
In the next section, we illustrate our approach by considering a non-trivial
example, the two-loop five-point nonplanar double pentagon diagram. This includes the IBP generation via the module intersection method, the massively parallel reduction of the IBP system and the basis conversion.

\section{The two-loop five-point nonplanar double pentagon example}\label{sec:doublepentagon}
In this section, we illustrate our IBP reduction method by applying it to a nontrivial example, 
the two-loop five-point nonplanar double pentagon. Note that a symbolic UT basis for this example was 
found in \cite{Abreu:2018aqd, Chicherin:2018yne}.  Furthermore, UT bases in terms of polylogarithm functions for the double pentagon and other 
two-loop five-point nonplanar massless integral families were analytically calculated in \cite{Chicherin:2018old}.

\begin{figure}
\centering
\begin{tikzpicture}[scale=0.8]
    \def\yys{0.5}
     \def\p{3}
    \draw [very thick] (0-\p/2,0+\yys) -- (3-\p/2,0+\yys)--(6-\p/2,0+\yys)--(6-\p/2,3+\yys)--(3-\p/2,3+\yys)--(0-\p/2,3+\yys)--(0-\p/2,0+\yys);
    \draw [very thick] (3-\p/2,0+\yys) -- (3-\p/2,3+\yys);
    \draw [very thick] (0-\p/2,0+\yys)--(-1-\p/2,-1+\yys);
    \draw [very thick] (6-\p/2,0+\yys)--(7-\p/2,-1+\yys);
    \draw [very thick] (6-\p/2,3+\yys)--(7-\p/2,4+\yys);
    \draw [very thick] (0-\p/2,3+\yys)--(-1-\p/2,4+\yys);
    \draw [very thick] (3-\p/2,1.5+\yys)--(4-\p/2.2,1.5+\yys);
    \node [below left=0 cm] at (-1-\p/2,-1+\yys)  {$p_1$};
    \node [above left=0 cm] at (-1-\p/2,4+\yys)   {$p_2$};
    \node [right=0 cm] at (4.2-\p/2,1.5+\yys) {$p_3$};
    \node [above right= 0 cm] at (7-\p/2,4+\yys) {$p_4$};
    \node [below right= 0 cm] at (7-\p/2,-1+\yys)  {$p_5$};
    \node [below=0 cm] at (1.5-\p/2,0+\yys)  {$z_1$};
    \node [left=0 cm] at (0-\p/2,1.5+\yys)  {$z_2$};
    \node [above=0 cm] at (1.5-\p/2,3+\yys)  {$z_3$};
    \node [below=0 cm] at (4.5-\p/2,0+\yys)  {$z_4$};
    \node [above=0 cm] at (4.5-\p/2,3+\yys)  {$z_5$};
    \node [right=0 cm] at (6-\p/2,1.5+\yys)  {$z_6$};
    \node [right=0 cm] at (3-\p/2,0.75+\yys)  {$z_7$};
    \node [right=0 cm] at (3-\p/2,2.25+\yys)  {$z_8$};
	\draw [dotted] (-4.5-\p,-1.5)--(10.5,-1.5)--(10.5,-11)--(-4.5-\p,-11)--(-4.5-\p,-1.5); 
    \def\ys{-0.5}
	\draw [very thick] (-3-\p,-3)--(-3-\p,-5);
		\draw [very thick] (-3-\p,-4) ellipse (0.6cm and 1cm);
			\draw [very thick] (-3-\p,-5)--(-2.55-\p,-5.5); 
			\draw [very thick] (-3-\p,-5)--(-3.45-\p,-5.5); 
			\draw [very thick] (-3-\p,-5)--(-3-\p,-5.65);
			\draw [very thick] (-3-\p,-3)--(-2.55-\p,-2.5); 
			\draw [very thick] (-3-\p,-3)--(-3.45-\p,-2.5);  
				\node [below=0 cm] at (-2.55-\p,-5.5)  {$p_2$};
    				\node [below=0 cm] at (-3.45-\p,-5.5)   {$p_1$};
    				\node [below=0 cm] at (-3-\p,-5.65) {$p_3$};
    				\node [above= 0 cm] at (-2.55-\p,-2.5) {$p_5$};
    				\node [above= 0 cm] at (-3.45-\p,-2.5)  {$p_4$};
    			\draw[line width=0.10cm,red] (-2.8-\p,-4)--(-3.2-\p,-4);
    			\draw[line width=0.10cm,red] (-2.2-\p,-4)--(-2.6-\p,-4);
    			\draw[line width=0.10cm,red] (-3.4-\p,-4)--(-3.8-\p,-4);  
    				\node [above left=0 cm] at (-3-\p,-4)  {$7$};
    				\node [above left=0 cm] at (-2.4-\p,-4)   {$5$};
    				\node [above left=0 cm] at (-3.6-\p,-4) {$1$}; 
	\draw [very thick] (-0-\p,-3)--(-0-\p,-5);
		\draw [very thick] (0-\p,-4) ellipse (0.6cm and 1cm);
			\draw [very thick] (-3,-5)--(-2.55,-5.5); 
			\draw [very thick] (-3,-5)--(-3.45,-5.5); 
			\draw [very thick] (-3,-3)--(-2.55,-2.5); 
			\draw [very thick] (-3,-3)--(-3.45,-2.5); 
			\draw [very thick] (-3,-3)--(-3,-2.35); 
				\node [below=0 cm] at (-2.55,-5.5)  {$p_2$};
    				\node [below=0 cm] at (-3.45,-5.5)  {$p_1$};
    				\node [above=0 cm] at (-2.55,-2.5) {$p_5$};
    				\node [above= 0 cm] at (-3.45,-2.5) {$p_3$};
    				\node [above= 0 cm] at (-3,-2.35)  {$p_4$};
    			\draw[line width=0.10cm,red] (-2.8,-4)--(-3.2,-4);
    			\draw[line width=0.10cm,red] (-2.2,-4)--(-2.6,-4);
    			\draw[line width=0.10cm,red] (-3.4,-4)--(-3.8,-4);  
    				\node [above left=0 cm] at (-3,-4)  {$8$};
    				\node [above left=0 cm] at (-2.4,-4)   {$5$};
    				\node [above left=0 cm] at (-3.6,-4) {$1$}; 
	\draw [very thick] (3-\p,-3)--(3-\p,-5);
		\draw [very thick] (3-\p,-4) ellipse (0.6cm and 1cm);
			\draw [very thick] (-3+\p,-5)--(-2.55+\p,-5.5); 
			\draw [very thick] (-3+\p,-5)--(-3.45+\p,-5.5); 
			\draw [very thick] (-3+\p,-5)--(-3+\p,-5.65);
			\draw [very thick] (-3+\p,-3)--(-2.55+\p,-2.5); 
			\draw [very thick] (-3+\p,-3)--(-3.45+\p,-2.5);  
				\node [below=0 cm] at (-2.55+\p,-5.5)  {$p_4$};
    				\node [below=0 cm] at (-3.45+\p,-5.5)   {$p_1$};
    				\node [below=0 cm] at (-3+\p,-5.65) {$p_2$};
    				\node [above= 0 cm] at (-2.55+\p,-2.5) {$p_5$};
    				\node [above= 0 cm] at (-3.45+\p,-2.5)  {$p_3$};
    			\draw[line width=0.10cm,red] (-2.8+\p,-4)--(-3.2+\p,-4);
    			\draw[line width=0.10cm,red] (-2.2+\p,-4)--(-2.6+\p,-4);
    			\draw[line width=0.10cm,red] (-3.4+\p,-4)--(-3.8+\p,-4);  
    				\node [above left=0 cm] at (-3+\p,-4)  {$8$};
    				\node [above left=0 cm] at (-2.4+\p,-4)   {$6$};
    				\node [above left=0 cm] at (-3.6+\p,-4) {$1$}; 
	\draw [very thick] (6-\p,-3)--(6-\p,-5);
		\draw [very thick] (6-\p,-4) ellipse (0.6cm and 1cm);
			\draw [very thick] (-3+2*\p,-5)--(-2.55+2*\p,-5.5); 
			\draw [very thick] (-3+2*\p,-5)--(-3.45+2*\p,-5.5); 
			\draw [very thick] (-3+2*\p,-3)--(-2.55+2*\p,-2.5); 
			\draw [very thick] (-3+2*\p,-3)--(-3.45+2*\p,-2.5); 	
			\draw [very thick] (-3+2*\p,-3)--(-3+2*\p,-2.35); 
				\node [below=0 cm] at (-2.55+2*\p,-5.5)  {$p_3$};
    				\node [below=0 cm] at (-3.45+2*\p,-5.5)  {$p_1$};
    				\node [above=0 cm] at (-2.55+2*\p,-2.5) {$p_2$};
    				\node [above= 0 cm] at (-3.45+2*\p,-2.5) {$p_5$};
    				\node [above= 0 cm] at (-3+2*\p,-2.35)  {$p_4$};
    			\draw[line width=0.10cm,red] (-2.8+2*\p,-4)--(-3.2+2*\p,-4);
    			\draw[line width=0.10cm,red] (-2.2+2*\p,-4)--(-2.6+2*\p,-4);
    			\draw[line width=0.10cm,red] (-3.4+2*\p,-4)--(-3.8+2*\p,-4);  
    				\node [above left=0 cm] at (-3+2*\p,-4)  {$8$};
    				\node [above left=0 cm] at (-2.4+2*\p,-4)   {$4$};
    				\node [above left=0 cm] at (-3.6+2*\p,-4) {$2$}; 	
	\draw [very thick] (9-\p,-3)--(9-\p,-5);
		\draw [very thick] (9-\p,-4) ellipse (0.6cm and 1cm);
			\draw [very thick] (-3+3*\p,-5)--(-2.55+3*\p,-5.5); 
			\draw [very thick] (-3+3*\p,-5)--(-3.45+3*\p,-5.5); 
			\draw [very thick] (-3+3*\p,-3)--(-2.55+3*\p,-2.5); 
			\draw [very thick] (-3+3*\p,-3)--(-3.45+3*\p,-2.5); 
			\draw [very thick] (-3+3*\p,-3)--(-3+3*\p,-2.35);  
				\node [below=0 cm] at (-2.55+3*\p,-5.5)  {$p_3$};
    				\node [below=0 cm] at (-3.45+3*\p,-5.5)   {$p_2$};
    				\node [above=0 cm] at (-2.55+3*\p,-2.5) {$p_5$};
    				\node [above= 0 cm] at (-3.45+3*\p,-2.5) {$p_1$};
    				\node [above= 0 cm] at (-3+3*\p,-2.35)  {$p_4$};
    			\draw[line width=0.10cm,red] (-2.8+3*\p,-4)--(-3.2+3*\p,-4);
    			\draw[line width=0.10cm,red] (-2.2+3*\p,-4)--(-2.6+3*\p,-4);
    			\draw[line width=0.10cm,red] (-3.4+3*\p,-4)--(-3.8+3*\p,-4);  
    				\node [above left=0 cm] at (-3+3*\p,-4)  {$7$};
    				\node [above left=0 cm] at (-2.4+3*\p,-4)   {$5$};
    				\node [above left=0 cm] at (-3.6+3*\p,-4) {$2$}; 	
	\draw [very thick] (9,-3)--(9,-5);
		\draw [very thick] (9,-4) ellipse (0.6cm and 1cm);
			\draw [very thick] (-3+4*\p,-5)--(-2.55+4*\p,-5.5); 
			\draw [very thick] (-3+4*\p,-5)--(-3.45+4*\p,-5.5); 
			\draw [very thick] (-3+4*\p,-3)--(-2.55+4*\p,-2.5); 
			\draw [very thick] (-3+4*\p,-3)--(-3.45+4*\p,-2.5); 
			\draw [very thick] (-3+4*\p,-3)--(-3+4*\p,-2.35);  
				\node [below=0 cm] at (-2.55+4*\p,-5.5)  {$p_5$};
    				\node [below=0 cm] at (-3.45+4*\p,-5.5)   {$p_1$};
    				\node [above=0 cm] at (-2.55+4*\p,-2.5) {$p_3$};
    				\node [above= 0 cm] at (-3.45+4*\p,-2.5) {$p_2$};
    				\node [above= 0 cm] at (-3+4*\p,-2.35)  {$p_4$};
    			\draw[line width=0.10cm,red] (-2.8+4*\p,-4)--(-3.2+4*\p,-4);
    			\draw[line width=0.10cm,red] (-2.2+4*\p,-4)--(-2.6+4*\p,-4);
    			\draw[line width=0.10cm,red] (-3.4+4*\p,-4)--(-3.8+4*\p,-4);  
    				\node [above left=0 cm] at (-3+4*\p,-4)  {$7$};
    				\node [above left=0 cm] at (-2.4+4*\p,-4)   {$6$};
    				\node [above left=0 cm] at (-3.6+4*\p,-4) {$2$}; 		
	\draw [very thick] (-3-\p,-7+\ys)--(-3-\p,-9+\ys);
		\draw [very thick] (-3-\p,-8+\ys) ellipse (0.6cm and 1cm);
			\draw [very thick] (-3-\p,-9+\ys)--(-2.55-\p,-9.5+\ys); 
			\draw [very thick] (-3-\p,-9+\ys)--(-3.45-\p,-9.5+\ys); 
			\draw [very thick] (-3-\p,-7+\ys)--(-2.55-\p,-6.5+\ys); 
			\draw [very thick] (-3-\p,-7+\ys)--(-3.45-\p,-6.5+\ys); 
			\draw [very thick] (-3-\p,-7+\ys)--(-3-\p,-6.35+\ys); 
				\node [below=0 cm] at (-2.55-\p,-9.5+\ys)  {$p_4$};
    				\node [below=0 cm] at (-3.45-\p,-9.5+\ys)   {$p_2$};
    				\node [above=0 cm] at (-2.55-\p,-6.5+\ys) {$p_1$};
    				\node [above= 0 cm] at (-3.45-\p,-6.5+\ys){$p_3$};
    				\node [above= 0 cm] at (-3-\p,-6.35+\ys)  {$p_5$};
    			\draw[line width=0.10cm,red] (-2.8-\p,-8+\ys)--(-3.2-\p,-8+\ys);
    			\draw[line width=0.10cm,red] (-2.2-\p,-8+\ys)--(-2.6-\p,-8+\ys);
    			\draw[line width=0.10cm,red] (-3.4-\p,-8+\ys)--(-3.8-\p,-8+\ys);  
    				\node [above left=0 cm] at (-3-\p,-8+\ys)  {$8$};
    				\node [above left=0 cm] at (-2.4-\p,-8+\ys)   {$6$};
    				\node [above left=0 cm] at (-3.6-\p,-8+\ys) {$2$};   	
	\draw [very thick] (-0-\p,-7+\ys)--(-0-\p,-9+\ys);
		\draw [very thick] (0-\p,-8+\ys) ellipse (0.6cm and 1cm);
			\draw [very thick] (-3,-9+\ys)--(-2.55,-9.5+\ys); 
			\draw [very thick] (-3,-9+\ys)--(-3.45,-9.5+\ys); 
			\draw [very thick] (-3,-7+\ys)--(-2.55,-6.5+\ys); 
			\draw [very thick] (-3,-7+\ys)--(-3.45,-6.5+\ys); 
			\draw [very thick] (-3,-7+\ys)--(-3,-6.35+\ys); 
				\node [below=0 cm] at (-2.55,-9.5+\ys)  {$p_2$};
    				\node [below=0 cm] at (-3.45,-9.5+\ys)  {$p_1$};
    				\node [above=0 cm] at (-2.55,-6.5+\ys) {$p_5$};
    				\node [above= 0 cm] at (-3.45,-6.5+\ys) {$p_3$};
    				\node [above= 0 cm] at (-3,-6.35+\ys)  {$p_4$};
    			\draw[line width=0.10cm,red] (-2.8,-8+\ys)--(-3.2,-8+\ys);
    			\draw[line width=0.10cm,red] (-2.2,-8+\ys)--(-2.6,-8+\ys);
    			\draw[line width=0.10cm,red] (-3.4,-8+\ys)--(-3.8,-8+\ys);  
    				\node [above left=0 cm] at (-3,-8+\ys)  {$7$};
    				\node [above left=0 cm] at (-2.4,-8+\ys)   {$4$};
    				\node [above left=0 cm] at (-3.6,-8+\ys) {$3$}; 
	\draw [very thick] (3-\p,-7+\ys)--(3-\p,-9+\ys);
		\draw [very thick] (3-\p,-8+\ys) ellipse (0.6cm and 1cm);
			\draw [very thick] (-3+\p,-9+\ys)--(-2.55+\p,-9.5+\ys); 
			\draw [very thick] (-3+\p,-9+\ys)--(-3.45+\p,-9.5+\ys); 
			\draw [very thick] (-3+\p,-9+\ys)--(-3+\p,-9.65+\ys);
			\draw [very thick] (-3+\p,-7+\ys)--(-2.55+\p,-6.5+\ys); 
			\draw [very thick] (-3+\p,-7+\ys)--(-3.45+\p,-6.5+\ys);  
				\node [below=0 cm] at (-2.55+\p,-9.5+\ys)  {$p_2$};
    				\node [below=0 cm] at (-3.45+\p,-9.5+\ys)   {$p_1$};
    				\node [below=0 cm] at (-3+\p,-9.65+\ys) {$p_3$};
    				\node [above= 0 cm] at (-2.55+\p,-6.5+\ys) {$p_5$};
    				\node [above= 0 cm] at (-3.45+\p,-6.5+\ys)  {$p_4$};
    			\draw[line width=0.10cm,red] (-2.8+\p,-8+\ys)--(-3.2+\p,-8+\ys);
    			\draw[line width=0.10cm,red] (-2.2+\p,-8+\ys)--(-2.6+\p,-8+\ys);
    			\draw[line width=0.10cm,red] (-3.4+\p,-8+\ys)--(-3.8+\p,-8+\ys);  
    				\node [above left=0 cm] at (-3+\p,-8+\ys)  {$8$};
    				\node [above left=0 cm] at (-2.4+\p,-8+\ys)   {$4$};
    				\node [above left=0 cm] at (-3.6+\p,-8+\ys) {$3$};     				
	\draw [very thick] (3,-7+\ys)--(3,-9+\ys);
		\draw [very thick] (3,-8+\ys) ellipse (0.6cm and 1cm);
			\draw [very thick] (-3+2*\p,-9+\ys)--(-2.55+2*\p,-9.5+\ys); 
			\draw [very thick] (-3+2*\p,-9+\ys)--(-3.45+2*\p,-9.5+\ys); 
			\draw [very thick] (-3+2*\p,-9+\ys)--(-3+2*\p,-9.65+\ys); 
			\draw [very thick] (-3+2*\p,-7+\ys)--(-2.55+2*\p,-6.5+\ys); 
			\draw [very thick] (-3+2*\p,-7+\ys)--(-3.45+2*\p,-6.5+\ys); 
				\node [below=0 cm] at (-2.55+2*\p,-9.5+\ys)  {$p_5$};
    				\node [below=0 cm] at (-3.45+2*\p,-9.5+\ys)   {$p_1$};
    				\node [below=0 cm] at (-3+2*\p,-9.65+\ys) {$p_2$};
    				\node [above= 0 cm] at (-2.55+2*\p,-6.5+\ys) {$p_4$};
    				\node [above= 0 cm] at (-3.45+2*\p,-6.5+\ys)  {$p_3$};
    			\draw[line width=0.10cm,red] (-2.8+2*\p,-8+\ys)--(-3.2+2*\p,-8+\ys);
    			\draw[line width=0.10cm,red] (-2.2+2*\p,-8+\ys)--(-2.6+2*\p,-8+\ys);
    			\draw[line width=0.10cm,red] (-3.4+2*\p,-8+\ys)--(-3.8+2*\p,-8+\ys);  
    				\node [above left=0 cm] at (-3+2*\p,-8+\ys)  {$7$};
    				\node [above left=0 cm] at (-2.4+2*\p,-8+\ys)   {$6$};
    				\node [above left=0 cm] at (-3.6+2*\p,-8+\ys) {$3$};     				
    	\draw [very thick] (7.5,-8+\ys)--(7.5,-7+\ys);	
    		\draw [very thick] (6.75,-8+\ys) ellipse (0.75cm and 0.6cm);
    		\draw [very thick] (8.25,-8+\ys) ellipse (0.75cm and 0.6cm);	
    			\draw [very thick] (6,-8+\ys)--(5.25,-8+\ys+0.707*0.75); 
			\draw [very thick] (6,-8+\ys)--(5.25,-8+\ys-0.707*0.75); 
			\draw [very thick] (9,-8+\ys)--(9.75,-8+\ys+0.707*0.75); 
			\draw [very thick] (9,-8+\ys)--(9.75,-8+\ys-0.707*0.75);  
    				\node [left=0 cm] at (5.25,-8+\ys+0.707*0.75)   {$p_2$};
    				\node [left=0 cm] at (5.25,-8+\ys-0.707*0.75) {$p_1$};
    				\node [right= 0 cm] at (9.75,-8+\ys+0.707*0.75) {$p_4$};
    				\node [right= 0 cm] at (9.75,-8+\ys-0.707*0.75)  {$p_5$}; 
    				\node [above= 0 cm] at (7.5,-7+\ys)  {$p_3$}; 
    			\draw[line width=0.10cm,red] (6.75,-8.4+\ys)--(6.75,-8.8+\ys);
    			\draw[line width=0.10cm,red] (6.75,-7.6+\ys)--(6.75,-7.2+\ys);
    			\draw[line width=0.10cm,red] (8.25,-8.4+\ys)--(8.25,-8.8+\ys);
    			\draw[line width=0.10cm,red] (8.25,-7.6+\ys)--(8.25,-7.2+\ys);      			  
				\node [below=0 cm] at (6.75,-8.8+\ys)  {$1$};    				
    				\node [above=0 cm] at (6.75,-7.2+\ys)  {$3$};
    				\node [below=0 cm] at (8.25,-8.8+\ys)   {$4$};
    				\node [above=0 cm] at (8.25,-7.2+\ys)
                                {$5$};    				    				  	\end{tikzpicture}
\caption{We depict the two-loop five-point nonplanar double pentagon
  diagram, writing $z_i$ for the Baikov variables, which are equal to the inverse
  propagators. In particular, $z_1=l_1^2$ and $z_4=l_2^2$. We
  also draw the $11$ spanning cuts of this integral family. These 
  correspond to the non-collapsible master integrals, before using symmetries.}
  \label{fig dp}
\end{figure}
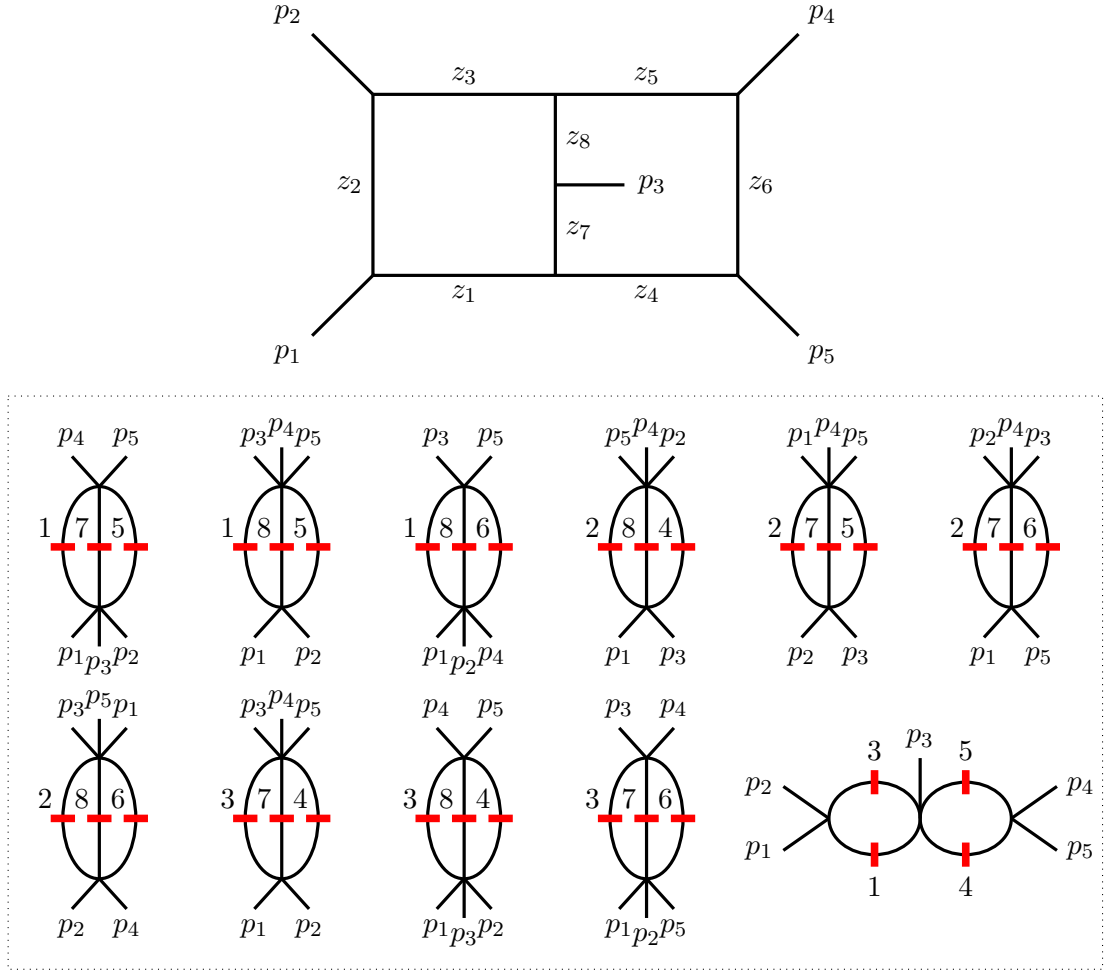

For the diagram in Figure \ref{fig dp}, we chose the following labeling for the propagators:
\begin{align}
D_1&= l_1^2 \qquad \qquad \quad D_2=(l_1-p_1)^2 \qquad   D_3=(l_1-p_{12})^2 \quad  D_4=l_2^2  \nonumber \\  D_5&=(l_2-p_{123})^2 \quad D_6=(l_2-p_{1234})^2 \quad D_7=(l_1-l_2)^2 \quad  D_8=(l_1-l_2+p_3)^2 \nonumber  \\  D_9&=(l_1-p_{1234})^2 \quad D_{10}=(l_2-p_1)^2 \quad D_{11}=(l_2-p_{12})^2 \,,
\end{align}
where the $l_i$ represent the loop momenta, the $p_i$ represent external
momenta,  and $p_{i \cdots j}= \sum_i^j p_i$. The first $8$ propagators represent the topology and the last three ones the irreducible scalar products.

This is a complicated integral family for IBP reduction, due to the number of independent scalars, which are s12, s23, s34, s45, s45, s15 and the spacetime dimension D, and due to the nonplanar topology with two pentagons inside. We demonstrate our method by reducing the $26$ integrals  listed in Figure  \ref{targets}
\begin{figure}
\begin{small}
\begin{gather}
  I[1,1,1,1,1,1,1,1,0,0,-4],I[1,1,1,1,1,1,1,1,0,-1,-3],I[1,1,1,1,1,1,1,1,0,-2,-2],\nonumber
  \\
I[1,1,1,1,1,1,1,1,0,-3,-1],I[1,1,1,1,1,1,1,1,0,-4,0],I[1,1,1,1,1,1,1,1,-1,0,-3],\nonumber\\
I[1,1,1,1,1,1,1,1,-1,-1,-2],I[1,1,1,1,1,1,1,1,-1,-2,-1],I[1,1,1,1,1,1,1,1,-1,-3,0],\nonumber\\
I[1,1,1,1,1,1,1,1,-2,0,-2],I[1,1,1,1,1,1,1,1,-2,-1,-1],I[1,1,1,1,1,1,1,1,-2,-2,0],\nonumber\\
I[1,1,1,1,1,1,1,1,-3,0,-1],I[1,1,1,1,1,1,1,1,-3,-1,0],I[1,1,1,1,1,1,1,1,-4,0,0],\nonumber\\
I[1,1,1,1,1,1,1,1,0,0,-3],I[1,1,1,1,1,1,1,1,0,-1,-2],I[1,1,1,1,1,1,1,1,0,-2,-1],\nonumber\\
I[1,1,1,1,1,1,1,1,0,-3,0],I[1,1,1,1,1,1,1,1,-1,0,-2],I[1,1,1,1,1,1,1,1,-1,-1,-1],\nonumber\\
I[1,1,1,1,1,1,1,1,-1,-2,0],I[1,1,1,1,1,1,1,1,-2,0,-1],I[1,1,1,1,1,1,1,1,-2,-1,0],\nonumber\\
I[1,1,1,1,1,1,1,1,-3,0,0],I[1,1,1,1,1,1,1,1,0,0,-2]\nonumber
\end{gather}\vspace{-5mm}
\end{small}
\caption{Integrals up to numerator degree $4$ without double propagators for the non-planar double pentagon diagram.}
\label{targets}
\end{figure}
to a master integral basis in the fashion of Laporta. Furthermore,
we convert the IBP coefficients to the coefficients of a dlog basis given
in \cite{Chicherin:2018old}. In this base change, we
observe a significant coefficient size reduction.

\subsection{Module intersection with cuts}
First, we use \textsc{Azurite} \cite{Georgoudis:2016wff} to find an
integral basis. Without considering symmetries, there are $113$
irreducible integrals, and with symmetries, there are $108$ master
integrals. Note that due to the number of master integrals, this IBP
reduction is significantly more complicated than the reduction of the
hexagon-box diagram in \cite{Boehm:2018fpv}, which has only $73$ master
integrals.

We can then construct the set of spanning cuts of this integral
family. Each spanning cut corresponds to a ``non-collapsible'' master
integral \cite{Georgoudis:2016wff}. There are $11$ spanning cuts
(without considering symmetries),
\begin{gather}
  \label{eq:11}
  \{1, 5, 7\}, \{1, 5, 8\}, \{1, 6, 8\}, \{2, 4, 8\}, \{2, 5, 7\},
  \{2, 6, 7\}, \nonumber \\
\{2, 6, 8\}, \{3, 4, 7\}, \{3, 4, 8\}, \{3, 6, 7\}, \{1, 3, 4, 5\},
\end{gather}
where the numbers indicate the propagators on cut. For example, $\{3,
4, 7\}$ means that $D_3 \mapsto 0, D_4 \mapsto 0, D_7 \mapsto 0$.

For each cut, we can apply our module intersection method to generate
IBPs without double propagators. In  \cite{Boehm:2018fpv}, the IBPs
are generated only from the top sector. Here, for each cut, we
generate the IBPs from both the top sector and lower sectors. For
example, for the cut $\{1,5,7\}$, we consider the $32$ sectors
supported on cut $\{1,5,7\}$, and compute all the module
intersections. This approach will generate more IBPs but make the IBP
system more block-triangular and easier for linear reduction.

With the {\bf degbound} option in {\sc Singular}, it is easy to generate all the module
intersections. For this integral family, choosing the degree bound $5$, and using
one core for each cut, it takes
less than $5$ minutes in total to solve all the module intersection problems
{\it analytically}. Later on, by finite-field methods, we find that with this choice of
degree bound, we obtain sufficiently many  IBPs for our problem.

After generating the IBPs, we use the two-step trimming process
described in Section~\ref{linear_reduction} to select necessary IBPs
for our targets. This computation is via finite-field methods and
powered by the package {\sc SpaSM}.

Note that different cuts can support the same master integral. We then speak of a cut overlap. For example, the integral $I[1, 1, 1, 1, 1, 0, 1, 1, 0, 0, 0]$
is supported by both cuts $\{1,5,7\}$ and $\{2,4,8\}$. The IBP reductions on these
two cuts should give the same coefficients. To avoid redundant
computations, we can apply the master integral removal method
\cite{Chawdhry:2018awn}, setting $I[1, 1, 1, 1, 1, 0, 1, 1, 0, 0, 0]\mapsto
0$ either in cut $\{1,5,7\}$ or in cut $\{2,4,8\}$. Clearly there are many
different removal strategies for master integral overlapping in
different cuts, and different strategies result in different
performances. In our computational problem, we find that the cuts
$\{1,6,8\}$ and $\{2,4,8\}$ are relatively ``difficult'' cuts for IBP
reduction. Hence, we set as many master integrals as possible in $\{1,6,8\}$
and $\{2,4,8\}$ to zero, and later on recover the remaining master integral
coefficients from other cuts via cut overlap.

We compute the module intersections analytically. For the purpose of
linear reduction, we further set
\begin{equation}
  \label{eq:17}
  s_{12}\mapsto 1, \quad c_2 \equiv s_{23}/s_{12}, \quad c_3 \equiv
  s_{34}/s_{12}, \quad c_4 \equiv s_{45}/s_{12}, \quad c_5 \equiv s_{15}/s_{12}
\end{equation}
to dehomogenize the IBP relations and speed up the computation. The
$s_{12}$ dependence can be recovered in the final step.

The resulting IBPs are summarized in
Table \ref{cut_IBP}. Note that for the cut $\{1,6,8\}$, there are $1203$
independent relations and $1205$ integrals after applying the idea of
\cite{Chawdhry:2018awn} to set most master integrals supported on the cut
$\{1,6,8\}$ to zero. As a result we only have to compute just two master integral
coefficients.

\begin{table}[ht]
  \centering
  \begin{tabular}{c|c|c|c|c|c}
     Cut & \# relations & \# integrals & size & $d_2$ & $d_3$\\
\hline
\{1,5,7\} &1134 & 1182& 0.77 MB & 21 &22 \\
\{1,5,8\} &1141 & 1192& 0.85 MB & 18 &18 \\
\{1,6,8\} &1203 & 1205& 1.1 MB & 19 & 30 \\
\{2,4,8\} &1245 & 1247& 1.1 MB & 35 & 24 \\
\{2,5,7\} &1164 & 1211& 0.84 MB & 26 & 18 \\
\{2,6,7\} &1147 & 1206 & 0.62 MB & 16 & 17 \\
\{2,6,8\} &1126 & 1177 & 0.83 MB & 16 & 18 \\
\{3,4,7\} &1172 & 1221 & 0.78 MB & 19 & 18 \\
\{3,4,8\} &1180 & 1226 & 1.0 MB & 19 & 22 \\
\{3,6,7\} &1115 & 1165 & 0.82 MB & 21 & 28 \\
\{1,3,4,5\} & 721 & 762 & 0.43MB & 14 & 14
  \end{tabular}
  \caption{The IBP relations generated on each cut by the module intersection
    method. We used finite-field methods to pick linearly independent and
     necessary IBP relations to reduce all target integrals. The size is the output file size
    on disk before reduction. The numbers $d_2$ and $d_3$ are the maximal degrees
    in the reduced IBP relations for $c_2$ and $c_3$, respectively. }
  \label{cut_IBP}
\end{table}

\subsection{IBP reduction}
We apply our reduction method via {\sc Singular} and \textsc{GPI-Space} to reduce
the linear systems in Table \ref{cut_IBP}. We use a semi-numeric
approach, choosing $c_4$, $c_5$ and the space-time dimension $D$ to be symbolic, and compute the
linear reduction with integer-valued $c_2$ and $c_3$.

By a linear reduction with $c_2$ (respectively $c_3$) symbolic and all the other parameters
numeric, we easily determine the maximal degree of $c_2$ (respectively $c_3$) in
the reduced IBP relations. The degrees are listed in  Table
\ref{cut_IBP} as $d_2$ and $d_3$, respectively. From this information, we get the minimal number
$(d_2+1)\times (d_3+1)$ of
semi-numeric computations for interpolating the analytic reduction
result. For example, for the cut $\{1,5,7\}$, we need to run
semi-numeric computations at least $506$ times.
Of course, the cuts exhibit different running times when performing the reductions: For instance, cut $\{1, 3, 4, 5\}$, which we already considered as an example in Section~\ref{sec:timing}, is the easiest in terms of running time, taking only about $11$ minutes when using $384$ CPU cores. In contrast, the cut $\{3,4,8\}$ is much more complex: its reduction took $12$ hours and $21$ minutes, using $384$ cores.

After getting the analytic reduction of all the cuts, we merge them to
get the full IBP reduction to a $113$-integral basis. Furthermore, we
apply the symmetries, obtained from~\cite{Georgoudis:2016wff},
\begin{align}
  \label{eq:18}
 I[0, 0, 1, 1, 0, 0, 0, 1, 0, 0, 0] &= I[1, 0, 0, 0, 1, 0, 1, 0,0,0, 0]\\
  I[0, 0, 1, 1, 0, 0, 1, 0, 0, 0, 0] &= I[1, 0, 0, 0, 1, 0, 0, 1,0,0, 0]\\
I[1, 0, 1, 0, 1, 0, 1, 0, 0, 0, 0] &= I[1, 0, 1, 1, 0, 0, 0, 1, 0,0,
                                     0]\\
I[0, 0, 1, 1, 1, 0, 1, 0, 0, 0, 0]&=I[1, 0, 0, 1, 1, 0, 0, 1, 0,0, 0]\\
I[1, 0, 1, 1, 1, 0, 0, 1, 0, 0, 0]&=I[1, 0, 1, 1, 1, 0, 1, 0, 0,
   0, 0]
\end{align}
to reduce the $26$ target integrals to a $108$-integral Laporta basis
$I$. We note
that the resulting file is large, with a size of $\sim 2.0$ GB on 
disk.

By setting all Mandelstam variables to integers, we have verified that our
result is consistent with FIRE6 \cite{Smirnov:2019qkx}.

\subsection{IBP coefficient conversion to a dlog basis}
In this subsection, we discuss converting the IBP coefficients for the Laporta
basis to the IBP coefficients of the dlog basis found in
\cite{Chicherin:2018old}. 

For this conversion, we again use the semi-numeric approach,
taking integer-valued $c_2$, $c_3$, and symbolic $c_4$,
$c_5$ and $D$, converting the coefficients and then interpolating. It is easy
to determine that the coefficients in the dlog basis have the
following maximal degrees for $c_2$ and $c_3$, respectively,
\begin{equation}
d_2' =20,\quad d_3'=20.
\end{equation}
By comparing with
Table  \ref{cut_IBP}, where $d_2$ can be as high as $35$, we find that the maximal degree drops. For the basis conversion, we carry
out a semi-numeric matrix multiplication with subsequent interpolation using
\textsc{Singular} and \textsc{GPI-Space}.

After the computation, we see that the IBP reduction coefficients of Figure \ref{targets} in this dlog basis
have size $480$ MB on disk, which shows a significant $76\%$
reduction of the IBP coefficient size compared to what we have for the
Laporta basis. On the other hand, if only the IBP reduction
coefficients in the dlog basis are needed, we can skip the interpolation for
the Laporta basis IBP coefficients, and directly convert the intermediate numerical
results to dlog basis IBP coefficients. Because of the maximal degree
drop, this shortcut reduces the required number of semi-numeric computations.

For convenience, we also provide the IBP coefficients in the dlog
basis, with the $s_{12}$ scalar recovered. All these analytic results
can be obtained via the links presented in the introduction of this
paper. Note that all files provided under the links contain $26\times 108$
matrices. For each matrix, the entry in the $i$th row and $j$th column is the corresponding IBP
coefficient for the $i$th target integral in Figure \ref{targets}, expanded
on the $j$th master integral. The Laporta basis and the dlog basis
are included in the auxiliary files of this paper. 

\section{Summary}
In this paper, we present our powerful new IBP reduction method, which is
based on computational algebraic geometry powered by the computer algebra system {\sc Singular} in conjunction with the taskflow management system \textsc{GPI-Space}. Our method is suitable for large scale IBP reduction
problems with complicated Feynman diagrams and multiple variables. We
demonstrate the power of the new method by  the analytic two-loop five-point
nonplanar double pentagon IBP computation. The computational result
has been cross-checked numerically using state-of-the-art IBP programs.

Our method is flexible and can be adapted in various different
scenarios: 
\begin{enumerate}
\item Modern methods for amplitude computation often follow the approach of numerically or semi-numerically calculating the IBP relations
in order to interpolate the
amplitude coefficient under consideration directly, instead of interpolating the analytic 
IBP relations. Our method can efficiently
compute the reduced numeric or semi-numeric IBP relations and, hence, perfectly fits
into this purpose. 
\item Our module intersection method can also be used for
integrals with double propagators or multiple-power propagators since
this IBP generating method avoids the increase of propagator exponents
and significantly reduces the size of the IBP system.
\item Although
our method is currently based on semi-numerical parallelizations
with integer-valued numerics, it clearly can be
extended to finite-field linear reduction, if necessary. 
\item More generally, our linear reduction parallelization method can
  be used for computational problems other than IBP
  reduction. For example, in recent years, it was found that the Bethe
  Ansatz equation of integrable spin chains can be analytically
  computed by algebraic geometry methods \cite{Jacobsen:2018pjt,
    Jiang:2017phk}. Often, this involves large-scale linear
  algebra computations with symbolic parameters, and our
  parallelization via the \textsc{Singular}-\textsc{GPI-Space} framework can greatly speed up the
  computation. We also expect that our reduction method can
  be used more generally for Gr\"obner basis computations with parameters.
\end{enumerate}

In the future, we will develop our code into an automated software package,
powered by \textsc{Singular} and \textsc{GPI-Space}, for solving large-scale IBP
or amplitude problems. The possible simplification of IBP coefficients
in a UT/dlog basis will be further investigated. We expect that this method will provide a boost for the
current NNLO precision computations.

\section*{Acknowledgments}
YZ thanks Johann Usovitsch for many enlightening discussions on IBP
reduction algorithms and their implementation, and Johannes Henn for
cutting-edge techniques for finding uniformly transcendental and dlog integrals.
We acknowledge Kasper Larsen for his work in the starting stage of
the analysis of the two-loop nonplanar five-point double pentagon
diagram. We thank Simon Badger, Rutger Boels, Christian Bogner, Hjalte
Frellesvig, Gudrun Heinrich, Harald Ita, Stephan Jahn, David Kosower, Roman  Lee,
Hui Luo, Yanqing Ma, Andreas von Manteuffel, Pierpaolo Mastrolia,
Alexander Mitov, Erik Panzer, Tiziano Peraro, Robert
Schabinger, Hans Sch\"onemann, Alexander Smirnov, Vladimir Smirnov, Peter Uwer, Gang Yang and Mao Zeng for their help on this
research project.

The research leading to these results has received funding from the Swiss
National Science Foundation (Ambizione grant PZ00P2 161341), from the
National Science Foundation under Grant No. NSF PHY17-48958, 
from the NSF of China with Grant No. 11235010, from
the European Research Council (ERC) under the European Union’s Horizon
2020 research and innovation programme (grant agreement No 725110).
The work of AG is supported by the Knut and Alice Wallenberg
Foundation under grant \#2015-0083. The work of DB, JB, and WD was
supported by Project II.5 of SFB-TRR 195 \emph{Symbolic Tools in
Mathematics and their Application} of the German Research Foundation
(DFG). The authors would also like to express a special thanks to the
Mainz Institute for Theoretical Physics (MITP) of the Cluster of
Excellence PRISMA+ (Project ID 39083149) for its hospitality and
support during the ``Mathematics of Linear Relations between
Feynman Integrals'' workshop.

\appendix

\section{Rational function interpolation}\label{sec:appendix}
\label{sec:rational_func_interpolation}
In this appendix, we introduce our simple approach to rational
function interpolation. Although this algorithm is rather
straight-forward compared to other more involved techniques available 
(see, for example,  \cite{zippel,bentiwari,KLW}), we have found that it is more suitable for our setup. The idea is to heuristically convert a rational function interpolation
problem to a polynomial function interpolation problem.

We focus on a general computational process with symbolic variables $x_1,
\ldots x_k$ which would give
the final result as a rational function
\begin{equation}
  \frac{F(x_1,\ldots, x_k)}{G(x_1,\ldots, x_k)} \,.
\label{result}
\end{equation}
Here, $F$ and $G$ are integer valued polynomials with $\gcd(F,G)=1$. Suppose we are in a situation where it is difficult to perform
this process with all parameters $x_1, \ldots, x_k$ symbolic, but where it is feasible to obtain results when some of the parameters are set
to be general integer values, and only the remaining parameters are symbolic. Then our approach is
to repeat such semi-numeric computations sufficiently often, and interpolate
to get \eqref{result}.

To be
specific, in such a situation, and for a fixed $k_1<k$, we refer to the computation with $x_{k_1+1},
\ldots, x_{k}$ symbolic, and the other parameters substituted by random integers,
\begin{equation}
  \label{evaluation}
   x_{1} \mapsto a_{1}^{(i)},\quad  \ldots, \quad x_{k_1+1}\mapsto a_{k_1+1}^{(i)},
\end{equation}
as the {\it{$i$th semi-numeric computation}}. We write the result of the $i$th semi-numeric computation as
\begin{equation}
  \label{eq:5}
  \frac{f_i(x_{k_1+1},
\ldots, x_{k})}{g_i(x_{k_1+1},
\ldots, x_{k})}\,.
\end{equation}
Note that, although computer algebra software can cancel the fraction
to get polynomials $f_i$ and $g_i$ with $\gcd(f_i,g_i)=1$, the relation between $G$ and $g_i$ is
not clear a priori. For example, it may happen that
\begin{equation}
  \label{eq:6a}
  g_i(x_{k_1+1},\ldots, x_{k}) \not = G(a_1^{(i)},\ldots, a_k^{(i)}, x_{k_1+1},\ldots, x_{k}).
\end{equation}
Similarly, we may have that
\begin{equation}
  \label{eq:6b}
  f_i(x_{k_1+1},\ldots, x_{k}) \not =F(a_1^{(i)}, \ldots, a_k^{(i)}, x_{k_1+1},\ldots, x_{k}).
\end{equation}
The reason for this is that after taking integer values $a_1^{(i)},\ldots,
a_k^{(i)}$ for the first $k$ variables,
there can be additional cancellations between
$F$ and $G$. This phenomenon makes the direct polynomial interpolation
of the $g_i$ and $f_i$ inapplicable.

We solve this cancellation problem in a heuristic way. 
\begin{enumerate}
\item First, we compute a ``reference'' result with symbolic $x_1,
  \ldots, x_k$ and random integer values for the other parameters, $ x_{k_1+1} \mapsto b_{k_1+1},\ldots, x_{k}\mapsto
  b_{k}$:
  \begin{equation}
    \label{eq:7}
    \frac{p(x_{1},
\ldots, x_{k_1})}{q(x_{1},
\ldots, x_{k_1})}\,,
  \end{equation}
where $p$ and $q$ are integer valued polynomials with
$\gcd(p,q)=1$. Generally, except for a statistically very small set of
points $b=(b_{k_1+1},\ldots, b_{k})$, we can assume that the two polynomials $F(x_1, \ldots, x_{k_1}, b_{k_1+1},\ldots, b_{k})$
and $G(x_1, \ldots, x_{k_1}, b_{k_1+1},\ldots, b_{k})$ are coprime. Then, since
\begin{equation}
  \label{eq:8}
  \frac{p(x_{1},
\ldots, x_{k_1})}{q(x_{1},
\ldots, x_{k_1})}=\frac{F(x_1, \ldots, x_{k_1}, b_{k_1+1},\ldots, b_{k})}{G(x_1, \ldots, x_{k_1}, b_{k_1+1},\ldots, b_{k})}
\end{equation}
by the unique factorization domain (UFD) property,
\begin{eqnarray}
  \label{eq:9}
  G(x_1, \ldots, x_{k_1}, b_{k_1+1},\ldots, b_{k})&= \mu\cdot q(x_{1},
\ldots, x_{k_1})\,,\label{G-ref}\\
  F(x_1, \ldots, x_{k_1}, b_{k_1+1},\ldots, b_{k})&= \mu\cdot p(x_{1},
\ldots, x_{k_1})\,,
\end{eqnarray}
for some integer $\mu$. Note that our algorithm relies only on the existence of such a $\mu$.

\item Second, we do a ``majority vote'' selection from among the semi-numeric results: For each of the polynomials, we record the leading exponent in $x_{k+1}, \ldots, x_{k}$, and determine the most frequently occurring exponent $\mathbf r$. Then we drop all
  semi-numeric result whose exponent is not equal to $\mathbf r$. This step
  ensures that the gcd of $F(a_1^{(i)}, \ldots, a_k^{(i)},
  c_{k_1+1},\ldots, c_{k})$
and $G(a_1^{(i)}, \ldots, a_k^{(i)}, x_{k_1+1},\ldots, x_{k})$ is just
an integer instead of a non-constant polynomial in $x_{k_1+1},\ldots,
x_{k}$. Again by the UFD property,
\begin{eqnarray}
  \label{eq:10}
  G(a_1^{(i)},\ldots, a_k^{(i)}, x_{k_1+1},\ldots, x_{k}) &= \lambda_i\cdot g_i(x_{k_1+1},\ldots, x_{k}) \label{G-point}\,,\\
F(a_1^{(i)}, \ldots, a_k^{(i)}, x_{k_1+1},\ldots, x_{k}) &= \lambda_i\cdot f_i(x_{k_1+1},\ldots, x_{k})\,,
\end{eqnarray}
where each $\lambda_i$ is an integer. Setting  $x_{1} \mapsto a_{1}^{(i)},
\ldots, x_{k_1+1}\mapsto a_{k_1+1}^{(i)}$ in \eqref{G-ref} and $ x_{k_1+1} \mapsto b_{k_1+1},  \ldots, x_{k}\mapsto
  b_{k}$ in \eqref{G-point}, we determine that
  \begin{eqnarray}
    \lambda_i =\mu \frac{q(a_{1}^{(i)},\ldots, a_{k_1}^{(i)} )}{g_i(b_{k_1+1},\ldots, b_{k})}\,.
\label{interpolation-normal}
 \end{eqnarray}

3. Define
\begin{eqnarray}
  \label{eq:3}
  \tilde g_i(x_{k_1+1},\ldots, x_{k}) & \equiv\frac{\lambda_i}{\mu}
                                       g_i(x_{k_1+1},\ldots, x_{k}) = \frac{q(a_{1}^{(i)},\ldots, a_{k_1}^{(i)} )}{g_i(b_{k_1+1},\ldots, b_{k})}
                                       g_i(x_{k_1+1},\ldots, x_{k})\,,  \\
  \tilde f_i(x_{k_1+1},\ldots, x_{k}) & \equiv\frac{\lambda_i}{\mu}  f_i(x_{k_1+1},\ldots, x_{k}) =\frac{q(a_{1}^{(i)},\ldots, a_{k_1}^{(i)} )}{g_i(b_{k_1+1},\ldots, b_{k})}
                                       f_i(x_{k_1+1},\ldots, x_{k})\,.
\end{eqnarray}
Then interpolate both polynomials $\tilde f_i(x_{k_1+1},\ldots,
x_{k})$ and $\tilde g_i(x_{k_1+1},\ldots, x_{k})$ for the selected
semi-numeric points by standard polynomial interpolation
algorithms, say by using Newton polynomials. From \eqref{interpolation-normal},
we see that the resulting two polynomials are
\begin{gather}
  \label{eq:4}
  \frac{1}{\mu}F(x_1,\ldots, x_k, x_{k_1+1},\ldots, x_{k}) ,\quad
  \frac{1}{\mu}G(x_1,\ldots, x_k, x_{k_1+1},\ldots, x_{k}) \,,
\end{gather}
so that the ratio of the polynomials gives the desired fraction $F/G$, while the integer factor $\mu$ is canceled out.

So with the compensation factors $\lambda_i/\mu$, we can use simple
polynomial interpolation algorithms to interpolate rational functions at the
extra cost of computing only one additional reference point.  If the degree 
of $F$ and $G$ in $x_j$ in  is $d_j$ ($1\leq j\leq k_1$), for a general choice of the interpolation points, we need to compute the semi-numeric result
\begin{gather}
  \label{eq:12}
  (d_1+1) \times \ldots \times  (d_{k_1}+1)
\end{gather}
times, plus one computation for the reference point. 

The resulting algorithm is
implemented using \textsc{Singular} in conjunction with \textsc{GPI-Space}.

In practice, this algorithm is extendable in many ways. For example, instead of splitting the variables $x_1,\ldots x_k$ into
two groups $(x_1 ,\ldots , x_{k_1})$ and $(x_{k_1+1} ,\ldots , x_{k})$, we
can split the variables in more groups and use our algorithm
recursively. The algorithm can also be combined with finite field
reconstruction.

\end{enumerate}

\bibliography{IBP_GPIspace}
\end{document}